\documentclass[fleqn,usenatbib]{mnras}
\usepackage{newtxtext,newtxmath}
\usepackage[T1]{fontenc}
\usepackage{ae,aecompl}
\usepackage{graphicx}   
\usepackage{epsfig}
\usepackage{psfig}
\usepackage{color}
\usepackage{xcolor}
\usepackage{lastpage}

\def\gs{\mathrel{\raise0.35ex\hbox{$\scriptstyle >$}\kern-0.5em
\lower0.40ex\hbox{{$\scriptstyle \sim$}}}}
\def\ls{\mathrel{\raise0.35ex\hbox{$\scriptstyle <$}\kern-0.5em
\lower0.40ex\hbox{{$\scriptstyle \sim$}}}}

\makeatother

\date{Accepted ---. Received ---; in original form ---}

\pubyear{2026}

\title[{\rm AS2UDSx}: $S_{\rm 870\mu m}$\,$\sim$\,1\,mJy sources and spheroid formation]{Extending the ALMA survey of the SCUBA-2 CLS UDS field:\\
Tracing the obscured formation of spheroids across $z$\,$\sim$\,1--4}

\author[Smail et al.]{Ian Smail$^{1}$,  Steven Gillman$^{2,3}$, Ugn{\.e} Dudzevi{\v{c}}i{\={u}}t{\.{e}}$^{2,3}$ \& A.\,M.\ Swinbank$^{1}$ \\
  $^{1}$ Centre for Extragalactic Astronomy, Department of Physics, Durham University, South Road, Durham DH1 3LE, UK\\
  $^{2}$ Cosmic Dawn Center (DAWN), Denmark\\
  $^{3}$ DTU-Space, Elektrovej, Building 328, 2800 Kgs. Lyngby, Denmark\\
}

\begin{document}

\label{firstpage}

\pagerange{1--20}
\maketitle

\begin{abstract}
 We investigate the properties of 870-$\mu$m selected galaxies at $z$\,$\sim$\,1--4 with far-infrared luminosities of $L_{\rm IR}$\,$\sim$\,10$^{11}$--10$^{13}$\,L$_\odot$, encompassing systems that dominate obscured activity at the peak of cosmic star formation, to identify variations in star-formation processes as a function of dust mass and redshift. We revisit ALMA 870-$\mu$m continuum maps from the ALMA/SCUBA-2 UDS (AS2UDS) survey, lowering the source selection threshold from 4.3\,$\sigma$ to 3.1\,$\sigma$ to enlarge the sample with $S_{\rm 870\mu m}$\,$\sim$\,1\,mJy. To reduce contamination from noise peaks, we match submillimetre sources to a $K$-selected galaxy sample and apply cuts on photometric redshift and near-infrared $(H-K)$ colour. This yields 84 sources in our extended AS2UDS survey, AS2UDSx, with $S_{\rm 870\mu m}$\,$=$\,0.3--2.2\,mJy, doubling the sample at $S_{\rm 870\mu m}$\,$\sim$\,1\,mJy relative to the original study. Using this expanded sample, we find that  submillimetre galaxies with $S_{\rm 870\mu m}$\,$\sim$\,1\,mJy at $z$\,$\gs$\,2.5 share properties with brighter, more active populations, while those at $z$\,$\ls$\,2.5 are distinct, with lower gas fractions, shorter depletion times, and stellar morphologies from James Webb Space Telescope imaging that show less structured dust obscuration, resembling less-active field galaxies. This indicates a shift in the characteristics of 870-$\mu$m-selected galaxies at $S_{\rm 870\mu m}$\,$\sim$\,1\,mJy and $z$\,$\sim$\,2, likely driven by the stability of their gas discs. Brighter and higher-redshift galaxies can sustain dense, globally unstable discs through efficient  gas accretion, powering compact obscured starbursts. In contrast, fainter systems at $z$\,$\ls$\,2.5 lack this accretion, leading to more stable discs and more extended dust continuum emission. This suggests a natural division around $S_{\rm 870\mu m}\sim$\,1\,mJy and $z$\,$\sim$\,2: lower-redshift, fainter sources represent the most active secularly-driven extended star-forming discs, while similar and brighter,  higher-redshift submillimetre galaxies form a distinct population of compact starbursts within massive, unstable, gas-rich discs, consistent with progenitors of massive spheroids.
\end{abstract}


\section{Introduction}

Over the past $\sim$\,40 years, far-infrared  and submillimetre surveys have unveiled heavily dust-obscured star-forming galaxies and AGN out to $z$\,$\sim $\,10 (see \citealt{Hodge20} for a recent review). These sources are members of the luminous, ultraluminous or hyperluminous dust-obscured galaxy populations which evolve strongly with redshift, being relatively unimportant at $z$\,$\sim$\,0 \citep{Soifer91}, but 
potentially dominating the total star-formation rate (SFR) density at $z$\,$\sim$\,1--4 (e.g., \citealt{Bouwens20,Liu26}). The importance of luminous infrared galaxies (LIRGs, $L_{\rm IR}=$\,10$^{11}$--10$^{12}$\,L$_\odot$) for understanding the star-formation history of the Universe was first highlighted by \citet{Dole06} who  stacked Spitzer Space Telescope  data to show that  $\gs $\,70 per cent of the $\sim $\,200-$\mu$m peak in the cosmic far-infrared  background  arises from LIRGs at $z$\,$\gs $\,1. The same holds true for the  $\sim$\,1-mm cosmic background, where high-redshift ultraluminous infrared galaxies (ULIRGs,  $L_{\rm IR}$\,$=$\,10$^{12}$--10$^{13}$\,L$_\odot$, for a review see  \citealt[][]{Lonsdale06}), which overlap significantly with the  submillimetre galaxy population (e.g., \citealt{Hughes98, Chapman05}), provide less than half the emission, with the majority arising from LIRGs below the $S_{\rm 870\mu m}$\,$\sim$\,2-mJy confusion limit of single-dish sub-millimetre surveys (e.g., \citealt{Knudsen08,Chen13,Hsu24}).   Similarly, \citep{Dudzeviciute21} compared sample of  450-$\mu$m and 850-$\mu$m selected galaxies to probe respectively the LIRG and ULIRG regimes and concluded from their relative space densities at $z$\,$\sim$\,1--3 that LIRGs are the main obscured population at $z$\,$\sim$\,1--2 where the cosmic star-formation rate density peaks, while ULIRGs dominate at higher redshifts.

These observations clearly demonstrate the potential importance LIRGs (and ULIRGs) for galaxy evolution in the early Universe,  but how do these populations relate to the bulk of the ``normal'' star-forming galaxies?    At $z$\,$\ls$\,1 ULIRGs appear distinct to normal galaxies as the majority of ULIRGs and a reasonable fraction of LIRGs have stellar morphologies suggesting their intense obscured activity has been triggered by strong gravitational interactions in gas-rich  systems \citep[e.g.,]{Armus87,Melnick90,Murphy96,Farrah01, Veilleux02, Colina05}.   Is the  rapid increase in the average star-formation rate in  galaxies at $z$\,$\gg$\,0   and the increased  fraction of this activity that is obscured an indicator that mergers and interactions are increasingly important at higher redshifts?  Is there a similar dichotomy in high-redshift star-forming population between a less active, secularly-evolving population and more luminous systems, including bright submillimetre galaxies, that result from mergers and interactions  \citep[e.g.,][]{Kartaltepe12, Hung14}?  Or is there a continuity in properties, implying similar star-formation processes across all populations?

Samples of galaxies with far-infrared luminosities characteristic of ULIRGs, corresponding to star-formation rates of $\sim$\,100--1000\,M$_\odot$\,yr$^{-1}$ and submillimetre flux densities of $S_{\rm 870\mu m}$\,$\sim$\,2--10\,mJy, have been well studied as these sources are sufficiently bright that they can be detected using submillimetre or far-infrared bolometer cameras on  single-dish telescopes such as SCUBA-2 on the 15-m James Clerk Maxwell Telescope (JCMT) or PACS and SPIRE instruments on the Herschel Space Observatory \citep[e.g.,]{Eales10,Lutz11,Geach17,Simpson19}.   These highly-obscured galaxies can be precisely located using sub/millimetre interferometers such as the Atacama Large Millimetre Array (ALMA), yielding samples of $\sim$\,100--1000 sources whose multi-wavelength properties can then be characterised \citep{Hodge13,Hatsukade18,Stach19,Simpson20,GomezGuijarro22,Adscheid24,QuirosRojas24}.

Early attempts to use high-resolution near-infrared imaging from Hubble Space Telescope (HST)  \citep{Swinbank10b,Kartaltepe12,Aguirre13,Targett13,Chen15,Cowie18}, or dust continuum \citep{Engel10,Hodge16} to investigate the incidence of mergers in these systems suggested that they were frequent.  However, the former studies were sensitive to the effects of structured  obscuration in these very dusty galaxies, while the samples in the latter studies were typically small. More recent work has focused on using new high-resolution, near-infrared imaging from the James Webb Space Telescope (JWST) to investigate their morphological properties to attempt to assess the role of minor and major mergers in these systems \citep{Chen22,Cheng23,Gillman23,Gillman24,McKinney25,Bodansky25,Chan25}.   These studies are less sensitive to the influence of structured dust absorption than the earlier HST work and broadly agree that around half of bright submillimetre galaxies ($S_{\rm 870\mu m}$\,$\gs$\,2\,mJy) at $z$\,$\gs$\,1 show potential evidence of interactions, but that unlike the situation at $z$\,$\sim$\,0, the majority of these ULIRG-luminosity systems do  not appear to be major mergers.

To understand the triggering of the crucial LIRG population at $z$\,$\gs$\,1 and their star-formation properties we need  to select a large and robust sample of these sources, ideally using unbiased samples identified in the restframe far-infrared or submillimetre wavebands \citep[e.g.,][]{Dudzeviciute21}. However,  as \citet{Dole06} showed, LIRGs at $z$\,$\gs $\,1--2 have submillimetre flux densities at $\sim$\,1\,mm of $\ls $\,1--2\,mJy. The challenge of constructing statistically significant samples of sources with submillimetre flux densities around $S_{\rm 870\mu m}$\,$\sim$\,1\,mJy is that the surface densities of these sources are low enough that interferometric surveys are relatively inefficient (i.e., they spend a lot of valuable time integrating on regions of sky that are devoid of detectable sources), while these flux densities are below the limits of current bolometer cameras, primarily as a result of confusion due to the modest angular resolution of these systems.  Four approaches have therefore been employed to tackle this problem:

Firstly, gravitational lensing by clusters offers one way to increase the sensitivity and reduce the confusion in far-infrared or submillimetre maps and so reach below the blank-field confusion limit to  study LIRGs at $z$\,$\gs $\,1 (e.g., \citealt{Blain97}).   Magnifications  of 2--5\,$\times$ are typical for sources seen through the centres of massive clusters, and rarer strongly-lensed (multiply-imaged) sources can have magnifications  of 10--50\,$\times$ (e.g., \citealt{Swinbank10}).   Surveys for lensed submillimetre or far-infrared sources have been undertaken by several groups using single-dish facilities (e.g., \citealt{Smail97,Knudsen08,Chen13,Rawle16,Sun22}).  More recent work has used ALMA, for example \citet{Fujimoto24} catalogued  24 potentially background lensed 1.2-mm sources with estimated apparent $S_{\rm 870\mu m}$\,$\ls$\,1.5\,mJy and tapered SNR\,$\geq$\,4.5 from ALMA mapping of 33 massive clusters (the high significance cut is needed to reduce the false-positive rates due to the large number of independent beams in wide-field ALMA mosaic observations), while \citet{Cheng23} studied 16 high-significance sources from a heterogeneous mix of lensed and unlensed sources with JWST imaging and ALMA continuum detections at 870\,$\mu$m, 1.1\,mm or 3.3\,mm.

However, there are several significant challenges to exploiting lensing studies.  Firstly, the estimated amplification factors (which are essential to derive both the intrinsic fluxes and  estimated survey areas) are  sensitive to both the precise  positions for the most highly amplified sources near the caustics and the granularity of the lens models,  as well as the typically-unknown redshifts of these sources (which includes contamination from potential cluster members), e.g., \cite{Fujimoto24}.   Gravitational lensing by galaxies is even more prone to such biases (including differential amplification, \citealt{Serjeant12}) and so such samples do not provide a simple and bias-free probe of the high-redshift population.  In addition, lensing studies are fundamentally unsuited for measuring  clustering   -- a key indicator of the nature of this population -- due to the limited area of the  highly-magnified background sky and correspondingly small sample sizes in each field.  Equally, as there only a few amplified sources in each cluster such studies can't efficiently investigate the multi-wavelength properties,  as is possible by exploiting the investment of resources that has been poured into the best-studied extragalactic survey fields (e.g., COSMOS, ECDFS, UDS, etc.).    

Secondly, as \citet{Dole06} showed, stacking analysis can provide statistical limits on the luminosities of populations with LIRG-like luminosities selected from other wavebands (e.g., \citealt{Decarli14}).  However, these are unsuited to detailed study of individual sources or  to map the variation in  properties within a sample.  Moreover, the reliance on selection in another waveband means that they cannot provide a {\it complete} census of the LIRG population -- e.g., only around half of high-redshift ULIRGs are detected in deep Spitzer 24-$\mu$m or Very Large Array radio data that are typically used for pre-selection (e.g., \citealt{Hodge13}).

Next, ALMA blank field survey can provide the necessary sensitivity  \citep{Hatsukade18, GonzalezLopez20},  although to cover large areas  these become  observationally expensive  and so they tend to be limited to a few 10's arcmin$^2$ in size  (and if this is contiguous, they are then sensitive to cosmic variance).   These surveys typically yield small samples with correspondingly modest statistics and potentially significant cosmic variance uncertainties:   \citet{Hatsukade18}  catalogued
ten SNR\,$\geq$\,5 1.2-mm sources from the ALMA survey of GOODS-S field that have estimated $S_{\rm 870\mu m}$\,$\ls$\,1.5\,mJy (see also \citealt[][]{GomezGuijarro22}).

In contrast, studies exploiting the ALMA archive  such as A$^3$COSMOS can provide larger effective survey areas.  A$^3$COSMOS \citep{Adscheid24} has 23 blind-selected ALMA Band 7 ($\sim$\,870\,$\mu$m) continuum sources with $S_{\rm 870\mu m}$\,$\leq$\,1.5\,mJy and  SNR\,$\geq$\,5.4  that lie within the primary beam of ALMA archive observations in the COSMOS field, but not close to the phase centre (so are not the target of the heterogeneous science goals of the archival projects used to construct the catalogue).  At somewhat longer wavelengths there are another 19 ALMA Band 6 ($\sim$\,1150\,$\mu$m)  continuum sources in A$^3$COSMOS with the same SNR and beam selection and  1.1-mm flux densities that suggest $S_{\rm 870\mu m}$\,$\ls$\,1.5\,mJy.  However, even when detections of the primary science targets are excluded from these archival investigations, the remaining sources typically lie  in close proximity to (and potentially clustered around)  the targets of the diverse projects from which the original ALMA maps came.  This makes determing their true surface densities and understanding if they are an unbiased sample both complex and uncertain.

Finally, efficient ALMA surveys can be undertaken if these target the expected positions of faint submillimetre sources.   \citet{Hodge13} was an early example of such studies, mapping the ALMA counterparts in the vicinity of low-resolution 870-$\mu$m sources from a single-dish survey of the ECDFS field by \citet{Weiss09} and uncovering eight sources with $S_{\rm 870\mu m}$\,$\leq$\,1.5\,mJy  in their MAIN sample.  Similarly,  \citet{Cowie18} catalogued 12 870-$\mu$m ALMA detections with $S_{\rm 870\mu m}$\,$\leq$\,1.5\,mJy selected from follow up of  SCUBA-2 850-$\mu$m sources in the GOODS-S field (see also \citealt{McKay26}), while \citet{Stach19} identified 60  $S_{\rm 870\mu m}$\,$\leq$\,1.5\,mJy sources in similar ALMA follow-up of SCUBA-2 850-$\mu$m mapping of the $\sim$\,1-degree$^2$ UKIDSS UDS field.   These studies suffer some of the same issues as the compilations derived from the ALMA archive, such as the likelihood of clustering of new sources around the primary science targets and the difficulties of obtaining robust photometric constraints from any low-resolution wavebands as a result of brighter neighbouring sources.  However, the homogeneity of their selection means that in principle it is easier to assess any biases in the resulting samples.

In this work we applied the latter approach and extended the study of \citet{Stach19} to build a statistically reliable sample  of sources with $S_{\rm 870\mu m}$\,$\sim$\,1\,mJy, or $L_{\rm IR}$\,$\sim$\,10$^{12}$\,L$_\odot$,  around the LIRG/ULIRG boundary at $z$\,$\gs$\,1.    The AS2UDS survey \citep{Stach19} provides $\sim$\,50 arcmin$^{-2}$ of ALMA mapping at 870\,$\mu$m with $\sim$\,0.6$''$ resolution and typical 1-$\sigma$ sensitivity of 0.3\,mJy.   By probing  deeper in these regions it may be possible to  reach a depth where the population surface density was high enough to  mitigate some of the potential biases from  clustering of sources around the brighter target galaxies studied by AS2UDS.

%
%
\begin{figure*}
\centerline{
  \psfig{file=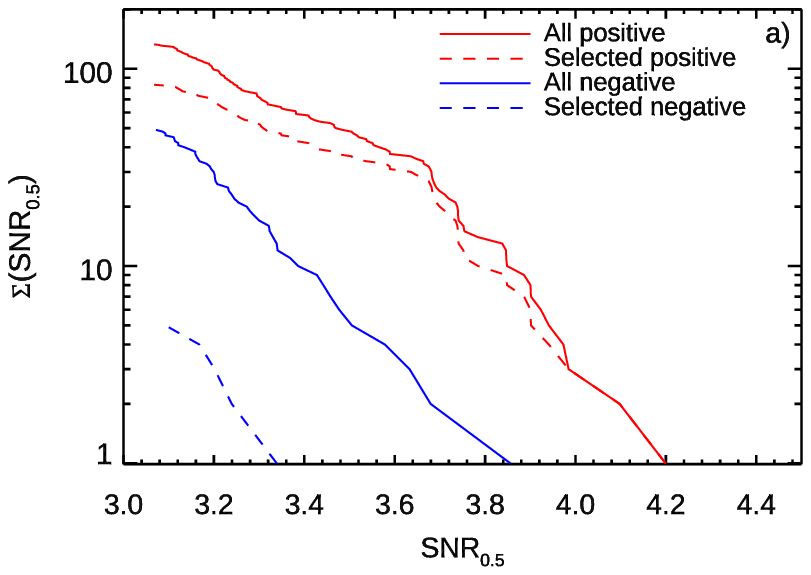, width=3.in, angle=0} 
  \psfig{file=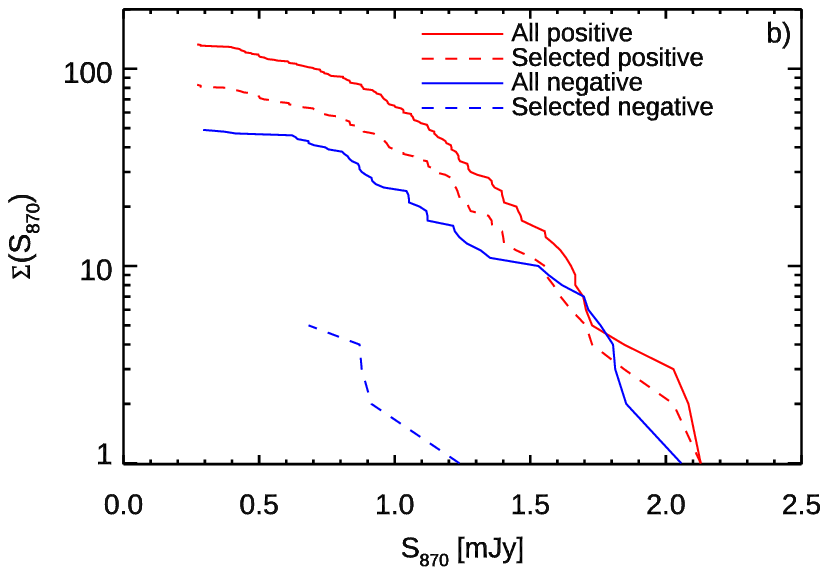, width=3.in, angle=0}}

\caption{\small
  The cumulative distributions of positive and (unphysical) negative sources matched to $K$-band counterparts (with likelihoods of being a random association of $P$\,$\leq$\,0.05)  as a function of  a) 870-$\mu$m signal-to-noise measured  in a 0.5$''$ diameter aperture, SNR$_{0.5}$, and b) 870-$\mu$m flux density,  $S_{\rm 870\mu m}$.   We also show the same distributions for the samples of positive and negative sources after applying selections for $z_{\rm med}$\,$\geq$\,1 and $(H-K)$\,$\geq$\,0.4,   as shown in Figure~2.   The selected positive sources have 870-$\mu$m flux densities ranging from $S_{\rm 870\mu m}$\,$\sim$\,0.27--2.2\,mJy.
}
\end{figure*}

The  aim of the \citet{Stach19} analysis of the ALMA maps taken for AS2UDS was to create a robust submillimetre-selected sample of bright sources across the $\sim$\,1-degree$^2$ UKIDSS UDS field.  As such they applied a high significance cut, $\geq$\,4.3\,$\sigma$, to identify the submillimetre galaxies in their ALMA maps to keep the contamination by spurious sources in their final catalogue to a $\ls$\,2 per cent false-positive rate.\footnote{The two per cent rate is for the whole sample and corresponds to $\sim$\,14 fake detections in the 707 catalogued sources.  The majority of these will have significances close to the 4.3-$\sigma$ cut and based on that we expect the AS2UDS sample to have $\sim$\,4 false sources from the 56 in the $S_{\rm 870\mu m}$\,$=$\,0.5--1.5\,mJy flux density range we focus on in our analysis below.  This corresponds to a seven per cent contamination rate and so we aim for a similar rate in our selection of AS2UDSx sources. }
They identified 707 submillimetre galaxies (excluding a bright strongly-lensed source, SXDF1100.001, \citealt[][]{Ikarashi11})  with flux densities that ranged from  $S_{\rm 870\mu m}$\,$\sim$\,0.6--11.9\,mJy across 716 870-$\mu$m ALMA maps,  $\sim$\,15 per cent  of the maps contained multiple sources and 101 maps showed no submillimetre source above 4.3\,$\sigma$.

For this faint extension of the original AS2UDS survey, hereafter AS2UDSx, our goal was to increase the numbers of the faintest sources, $S_{\rm 870\mu m}\ls$\,1.5--2\,mJy, detected in the ALMA maps by pushing to lower significance cuts and reducing the false positive rate by requiring a counterpart in the deep $K$-band imaging available for this field.  As we show, even this was not sufficient to reduce the false-positive  rate sufficiently at a detection significance  of $\gs$\,3\,$\sigma$ and we had to use the properties of the counterparts derived from the deep multi-wavelength coverage of this field to remove contamination from spurious  sources to achieve a final sample with a $\sim$\,7 per cent false-positive rate (see also \citealt{Chen16,An18,Franco20}).

This paper is structured as follows:   \S2  describes the construction of a deeper 870-$\mu$m source catalogue in the AS2UDS ALMA continuum maps, including the modelling of the spectral energy distributions of the candidate galaxies using {\sc magphys} and the use of their observed and inferred physical properties to reduce the contamination from spurious noise sources.    \S3 presents the analysis and  results relating to the  the properties of the faint sample and combines these with submillimetre galaxies with similar and brighter 870-$\mu$m flux densities from AS2UDS to trace the evolution in the properties of this population  as a function of dust mass and redshift.  The new sample of faint submillimetre galaxies allowed us to construct a larger sample of   $S_{\rm 870\mu m}$\,$\sim$\,1\,mJy systems with archival JWST NIRCam imaging, which provided stellar sizes and morphologies that were used to investigate the internal physical properties of these galaxies.   The results are discussed in \S4 and the main conclusions are given in \S5.

This study assumes a \cite{Chabrier03} IMF and a cosmology with $\Omega_{\rm M}$\,$=$\,0.3, $\Omega_\Lambda$\,$=$\,0.7 and $H_0$\,$=$\,70\,km\,s$^{-1}$\,Mpc$^{-1}$.     All quoted magnitudes are on the AB system and used 2.0$''$ diameter apertures with a point-source correction to total.  Errors on median values are estimated using bootstrap resampling.

\section{Observations, Reduction and Analysis}

\subsection{Cataloguing ALMA maps and matching to $K$-band counterparts}

We started the analysis by creating deeper catalogues of the 716 870-$\mu$m ALMA maps from the AS2UDS survey \citep{Stach19,Dudzeviciute20} of which 629 lie within the UDS $K$-band footprint (see below).   We used the 0.5$''$ tapered and cleaned 344-GHz maps ($\sim$\,870\,$\mu$m)  which were uncorrected for the primary beam sensitivity (the primary beam is 17.3$''$ diameter at 870\,$\mu$m) from \citet{Stach19}. These maps have a pixel scale of 0.06$''$ and typical noise levels of $\sim$\,0.3\,mJy, ranging from  $\sim$\,0.15--0.4\,mJy. More details of the data and their reduction can be found in \citet{Stach19}.

We ran {\sc SExtractor} on the original maps (denoted ``positive'' in the following) with a minimum detection area equal to the synthesised beam FWHM at a 2-$\sigma$ threshold above a local sky level, after  the maps were convolved with a 0.3$''$ FWHM gaussian filter. To assess the likely false-positive rate from noise in the maps we then inverted these maps and ran the identical search on the inverted (``negative'') maps.  We measured source positions, source signal-to-noise  in a 0.5$''$ diameter aperture, SNR$_{0.5}$, and flux densities in the same aperture.  The flux densities were converted to total assuming a point source correction and then corrected for primary beam attenuation and  for flux boosting using the simulations from \citet{Stach19}.   The  search of the original (positive) maps recovered all of the catalogued sources from the $\geq $\,4.3-$\sigma$ selected catalogue in \citet{Stach19} as well as many fainter features.   

To select sources for this study we adopted a   signal-to-noise limit of  SNR$_{0.5}$\,$\geq$\,3.1  motivated by the excess of sources in the ``positive'' maps relative to those in the ``negative'' maps above this threshold (see Figure~1).  We only retained sources that lie within the primary beam of their corresponding map and which were not included in the \citet{Stach19} catalogue.   Given the range of map sensitivities, the influence of the primary beam and the SNR range of our selection, we expected the sample of new faint sources to have 870-$\mu$m flux densities of $S_{\rm 870\mu m}$\,$\sim$\,0.3--2.5\,mJy, after accounting for possible primary beam corrections and deboosting,  with the majority having $S_{\rm 870\mu m}$\,$\sim$\,1\,mJy.

%
%
\begin{figure*}
  \raisebox{3.2ex}{
\begin{minipage}[h]{2.in}
    \centering
        \psfig{file=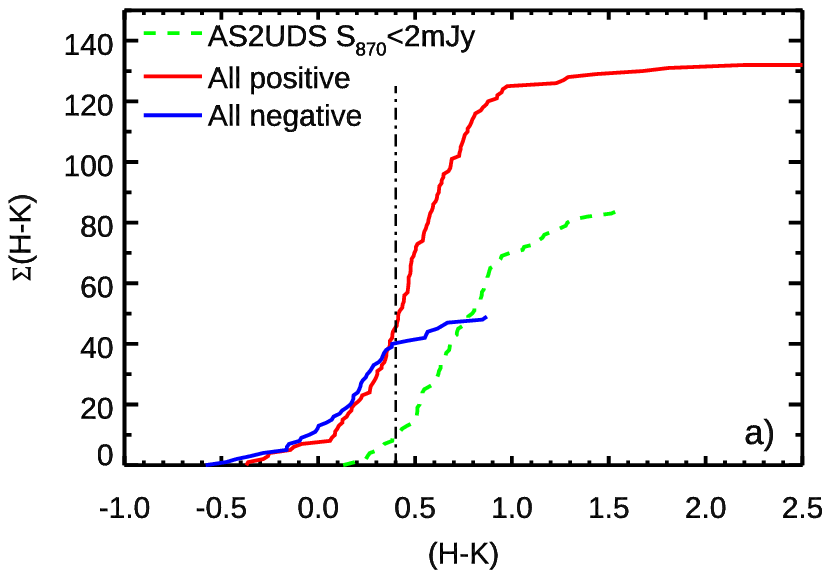, width=2.in, angle=0} 
        \psfig{file=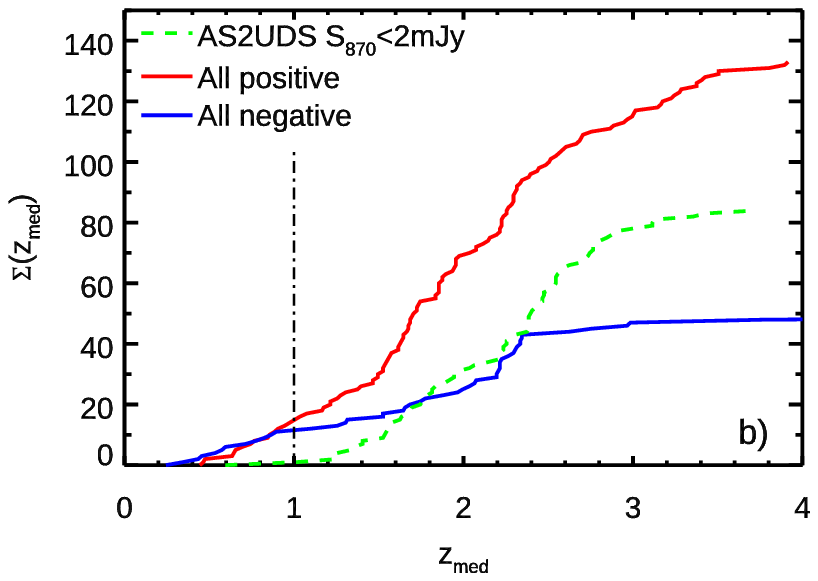, width=2.in, angle=0}
      \end{minipage}}
      \hspace*{-0.1in}\begin{minipage}[h]{4.7in}
    \psfig{file=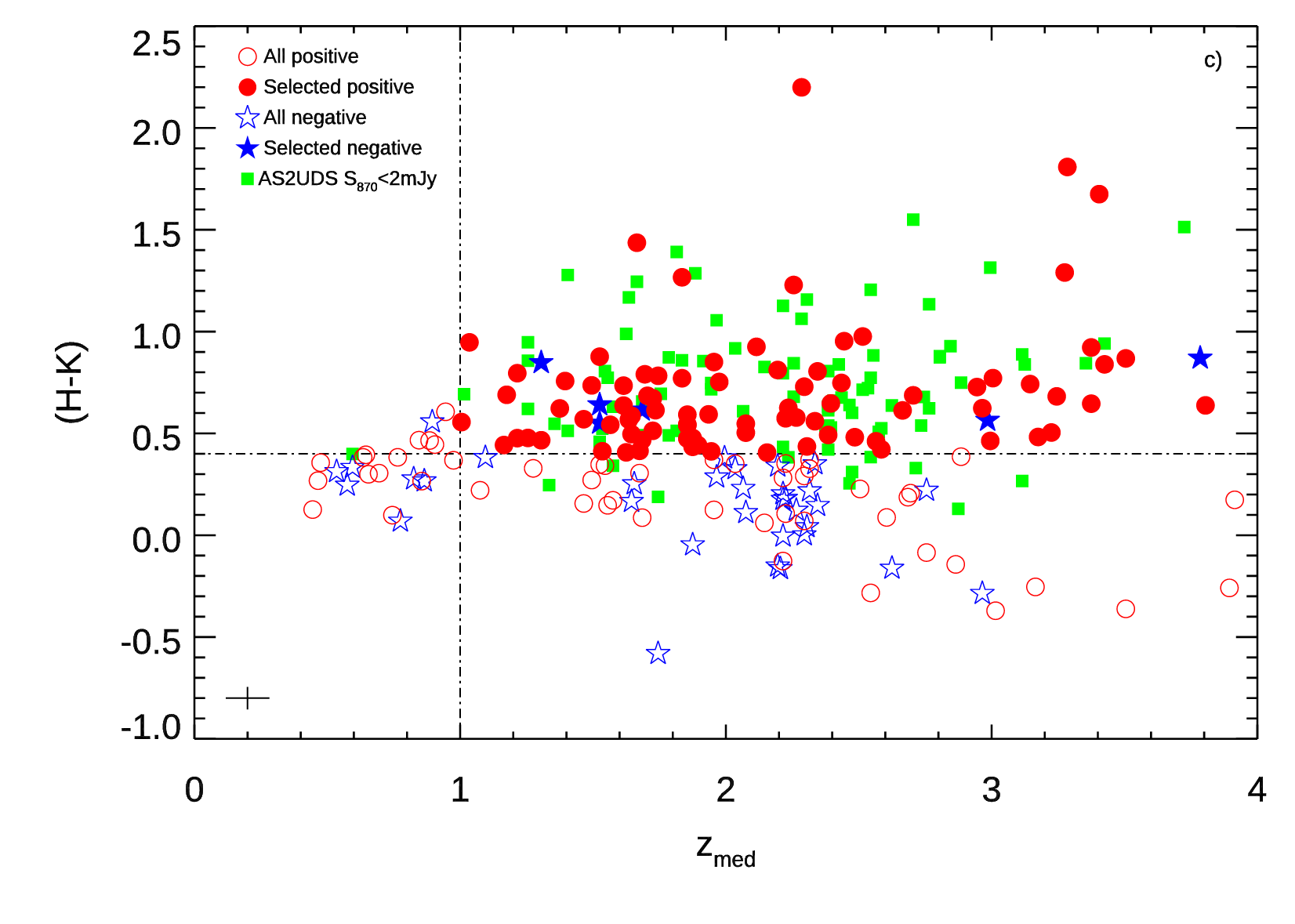, width=4.7in, angle=0}
  \end{minipage}
  \caption{\small
    a) The cumulative distribution of $K$-band-matched positive and negative sources as a function of $(H-K)$ colour. The numbers roughly match at $(H-K)$\,$\sim$\,0.4 (vertical line), beyond which the fraction of positive sources with red counterparts rises rapidly. The distribution for the $S_{\rm 870\mu m}$\,$\leq$\,2\,mJy AS2UDS sample \citep{Dudzeviciute20} shows the expected colours of real faint submillimetre galaxies, with most having $(H-K)$\,$\gs$\,0.4.
    b) The cumulative distribution of positive and negative sources as a function of {\sc magphys}-derived photometric redshift, $z_{\rm med}$. The AS2UDS $S_{\rm 870\mu m}$\,$\leq$\,2\,mJy sample illustrates that most real faint submillimetre galaxies have $z_{\rm med}$\,$\gs$\,1. The $z_{\rm med}$\,$\geq$\,1 cut corresponds to where positive and negative sources are comparable, above which positives rapidly dominate.
  c) The distribution of $(H-K)$ colour versus $z_{\rm med}$ for positive and negative samples, and the AS2UDS comparison sample. The adopted selection ($(H-K)$\,$\geq$\,0.4, $z_{\rm med}$\,$\geq$\,1; shown by lines) is $\sim$\,90 per cent complete based on AS2UDS, yielding 84 positive and 6 negative sources, implying $\sim$\,7 per cent contamination from spurious matches. Median errors for the positive sample are shown in the lower left.
  }
\end{figure*}

We next used the $K$-selected catalogue of sources derived from the very deep $K$-band coverage of the UDS region from the UKIDSS Ultra Deep Survey Data Release 11 \citep{Lawrence07}.  This catalogue comprises near-infrared and optical photometry of 296,007 sources for a $K$\,$\leq$\,25.7 selected sample using 2$''$ diameter apertures with point-source aperture corrections.\footnote{For a full description see:\\ https://www.nottingham.ac.uk/astronomy/UDS/DR11/}  These data have been supplemented by mid- and far-infrared and radio photometry, with a complete description of the multi-wavelength dataset  given by \citep{Dudzeviciute20}, we give a condensed summary below.

The near-infrared coverage includes UKIRT WFCAM \citep{Casali07} $JHK$ to 3-$\sigma$ depths of $J$\,$=$\,26.0, $H$\,$=$\,25.5 and $K$\,$=$\,25.7, with $Y$-band observations from VISTA/VIDEO \citep{Jarvis13} to $Y$\,$=$\,25.1.  The optical imaging includes deep Subaru Suprimecam  $BVRi'z'$-band with 3-$\sigma$ depths of 28.2, 27.6, 27.5, 27.5, and 26.4\,mag, respectively.  In addition, deep $U$-band photometry comes from the CFHT/Megacam survey and reaches a 3-$\sigma$ point-source depth of $U$\,$=$\,27.1\,mag.   These data were supplemented by photometry from Spitzer IRAC and MIPS covering 3.6--24\,$\mu$m, VLA radio data and Herschel SPIRE coverage at 250--500\,$\mu$m.  The latter data have coarse angular resolution and they were deblended using a prior catalogue comprising the bright ALMA sources, along with 24\,$\mu$m and 1.4\,GHz sources following the method described in \citet{Swinbank14}.    The multi-wavelength catalogue was astrometrically and photometrically calibrated as described in \citet{Dudzeviciute20}.

We matched the ALMA sources, both from the ``positive'' and inverted ``negative'' maps, to the $K$-band selected catalogue to identify potentially real faint submillimetre sources.  The matching used a 1$''$ search radius and we calculated the likelihood that any $K$-band sources found within this radius were chance matches from their $K$-band magnitude and radial offset, following \cite{Downes86} and \cite{Dunlop89}. $K$-band counterparts with probabilities of being random associations of less than $P$\,$\leq$\,0.05 were deemed to be ``reliable'' identifications and retained.

There were 1458 ``positive''  sources with an 870-$\mu$m SNR$_{0.5}$\,$\ge$\,3.1  within the primary beam areas of the  ALMA maps, of which 134 had an unique $K$-band counterpart within 1.0$''$ with a probability of being a random association of $P$\,$\leq$\,0.05 (Figure~1).   These ``positive'' sources had flux densities of $S_{\rm 870\mu m}$\,$\sim$\,0.27--2.2\,mJy as expected (Figure~1), and $K$-magnitudes of $K$\,$\sim$\,17.7--25.4.

The ``negative'' maps produced 1376 sources ($\sim$\,80 fewer than in the positive maps) within the primary beams with SNR$_{0.5}$\,$\geq$\,3.1, of which 50 had an unique $K$-band counterpart not assigned to another source (these had $K$\,$\sim$\,18.7--25.4), see Figure~1.   As the negative sources are unphysical, their matching rate to $K$-band counterparts provides an estimate of the contribution of false positive noise sources to our positive source catalogue.  The relative numbers indicated that roughly $\sim$\,80 of the 134 positive 3.1-$\sigma$ sources with $K$-band counterparts are likely to be real submillimetre galaxies, which suggested a contamination rate of $\sim$\,40 per cent.  While much reduced from the initial contamination rate in the sample, this was still too high to make use of the catalogue for our analysis and so we next had to reduce further the contamination using the properties of the matched $K$-band counterparts.

\subsection{Reducing contamination using physical properties}

Due to the distinctive multi-wavelength properties of submillimetre galaxies (see \citealt{Dudzeviciute20} and many previous studies, e.g., \citealt{Dannerbauer04,Frayer04,Pope05}), we  expected that the galaxy counterparts to the true positive submillimetre sources  would have characteristics that are distinct to those of  typical field galaxies which have been  matched to spurious noise peaks  in the catalogue.   The properties of the galaxy counterparts to the unphysical negative sources provided an indication of the  characteristics of this contamination and so we  used these  to reduce the contamination in the final catalogue.  To do this we modelled the multiband spectral energy distributions (SEDs) of  the positive and negative sources using {\sc magphys} \citep{daCunha08,daCunha15,Battisti19} following the same methodology as in \citet{Dudzeviciute20}.  A  summary is given here, but  the reader should refer to \citet{Dudzeviciute20} for a full description.

For this analysis, we used the updated high-redshift {\sc magphys} code from \cite{daCunha15} and \cite{Battisti19} that includes photometric redshift as a variable.  The code is optimised to fit the SEDs of high redshift ($z$\,$>$\,1) star-forming galaxies including modifications such as extended prior distributions for the star-formation history and dust optical depth effects, as well as the inclusion of absorption of UV photons by the foreground intergalactic medium.     To fit to the multi-wavelength photometry of our galaxies, {\sc magphys} generated a library of SEDs for a grid of redshifts for each star-formation history considered.  The code then selected the models that best-fit the multi-wavelength photometry by matching the model SEDs to the data using a $\chi^2$ test to yield the  best-fit parameters.  For our analysis we focused on seven of the derived  parameters: the photometric redshift, determined from the median of the posterior distribution ($z_{\rm med}$); star-formation rate (SFR), or equivalently far-infrared luminosity ($L_{\rm IR}$); stellar mass ($M_\ast$);   age; $V$-band attenuation ($A_{V}$)  and dust mass ($M_{\rm d}$).

We also analysed properties derived from the ratios of the physical parameters including specific star-formation rate, sSFR\,$=$\,${\rm SFR}/M_\ast$;  gas fraction, $M_{\rm gas}/M_\ast$; and gas depletion timescale, $T_{\rm dep}$\,$=$\,$M_{\rm gas}/{\rm SFR}$.  For the latter two properties we estimated the molecular gas mass, $M_{\rm gas}$, from the dust mass, $M_{\rm d}$ by adopting a gas-to-dust ratio, $\delta_{\rm GDR}$.  Various prescriptions have been proposed to estimate $\delta_{\rm GDR}$, including a simple constant value \citep{Swinbank14}, 
and more complex recipes employing assumptions about the gas-phase metallicity and its dependence on one or more of stellar mass, star-formation rate,  specific star-formation rate or redshift \citep[e.g.,][]{Magnelli12,GomezGuijarro22b,Tacconi20}.   In order of increasing complexity, the simplest prescriptions are those in \citet[][M12]{Magnelli12}, which is dependent on stellar mass and star-formation rate; and \citet[][GG22]{GomezGuijarro22b}, which uses stellar mass and redshift; while the prescription from \citet[][T20]{Tacconi20}  to derive $\delta_{\rm GDR}$ involves stellar mass,  star-formation rate, redshift and offset in specific star-formation rate from the ``main sequence''.

For the entire AS2UDSx+AS2UDS sample discussed later we estimated median (and 1-$\sigma$ dispersions) of $\delta_{\rm GDR}^{\rm M12}$\,$=$\,50\,$\pm$\,33,  $\delta_{\rm GDR}^{\rm GG22}$\,$=$\,100\,$\pm$\,64  and $\delta_{\rm GDR}^{\rm T20}$\,$=$\,123\,$\pm$\,231.  Focusing on the three flux-limited samples used in our analysis below,
$S_{\rm 870\mu m}$\,$=$\,0.5--1.5\,mJy, 2--4\,mJy and $\ge$\,4\,mJy, we estimated median $\delta_{\rm GDR}$
of $\delta_{\rm GDR}^{\rm M12}$\,$=$\,52 with a scatter of six per cent across the three samples,
$\delta_{\rm GDR}^{\rm GG22}$\,$=$\,100  with a scatter of four per cent, and 
$\delta_{\rm GDR}^{\rm T20}$\,$=$\,138 with a 52 per cent scatter.  
Apart from the systematic offset, the predicted $\delta_{\rm GDR}$ values tracked each other fairly closely for the \citet{GomezGuijarro22b}  and \citet{Magnelli12} estimates with a median ratio of  $\delta_{\rm GDR}^{\rm GG22}/\delta_{\rm GDR}^{M12}$\,$=$\,2.00\,$\pm$\,0.04 and a 1-$\sigma$ scatter of five per cent.  In contrast the \citet{Tacconi20} prescription showed a much larger dispersion with a $\sim$\,50 per cent scatter compared to either of the other two.  Given the limited information available on the gas-to-dust ratios or gas-phase metallicities of the submillimetre galaxy population across the flux and redshift range we probe \cite[][Taylor et al., in prep.]{Gillman26}, we  elected to simply adopt the $\delta_{\rm GDR}$ recipe from \citet{GomezGuijarro22b}  as these estimates typically lay in the middle of the range of choices.   We  also confirmed that adopting a fixed value of $\delta_{\rm GDR}$\,$=$\,100 following \citep{Swinbank14} did not qualitatively change the trends we found, beyond a modest systematic shift in the average estimated gas mass.

We modelled the spectral energy distributions of all 134 positive  and 50 negative sources using {\sc magphys}.  The fits used the available photometric detections or limits from the UDS photometric catalogue, including any Spitzer, Herschel or VLA detections and the flux densities measured from the ALMA maps (using the equivalent positive fluxes for the sources detected in the ``negative'' maps to replicate the behaviour of our modelling of the corresponding false sources in the ``positive'' sample).  We employed the calibrated photometry from \citet{Dudzeviciute20} that used  $\sim$\,6,700 galaxies with spectroscopic redshifts in the UDS field to determine median zero point offsets between the raw measured photometry and the predicted magnitudes of these galaxies at their spectroscopic redshift from their best-fit {\sc magphys} SEDs.  This removed any systematic offsets between the photometric system of the UDS observations and that assumed by {\sc magphys}.   Extensive testing of the derived {\sc magphys} properties are described in \citet{Dudzeviciute20}, here we note that from the photometric redshifts for the 11 sources in our sample with literature spectroscopic redshifts we derived a median $|z_{\rm med}-z_{\rm spec}|/(1+z_{\rm spec})$ of 0.07\,$\pm$\,0.05, lower than 0.13 scatter reported for the  AS2UDS submillimetre galaxies with spectroscopic redshifts by \citet{Dudzeviciute20}.

%
%
\setcounter{table}{1}
\begin{table*}
  \caption{Faint and bright samples. Median sample properties with bootstrap errors and 16$^{\rm th}$/84$^{\rm th}$ percentile ranges. }
  \setlength{\tabcolsep}{3pt}
\begin{tabular}{lcccccc}
  \hline \noalign {\smallskip}
  Property / $S_{\rm 870\mu m}$ & 0.5--1.5\,mJy & 0.5--1.5\,mJy & 0.5--1.5\,mJy & 2--4\,mJy &  $\geq$\,4\,mJy & $\geq$\,4\,mJy \\
  & & $z_{\rm med}$\,$=$\,1.5--2.5 & $z_{\rm med}$\,$=$\,2.5--3.5 & & & $K\leq 25.7$ \\
  \hline 
$N$ & 122 & 68 & 35 & 292 & 298 & 216 \\
  $S_{\rm 870\mu m}$ [mJy] &   1.13\,$\pm$\,0.04$_{-0.37}^{\, +0.27}$  & 1.15\,$\pm$\,0.06$_{-0.41}^{\, +0.26}$ & 1.10\,$\pm$\,0.07$_{-0.28}^{\, +0.11}$  & 3.09\,$\pm$\,0.06$_{-0.65}^{\, +0.72}$  & 5.34\,$\pm$\,0.12$_{-0.94}^{\, +1.72}$ & 5.24\,$\pm$\,0.14$_{-0.86}^{\, +1.67}$\\[1mm]
  $z_{\rm med}$ & 2.17\,$\pm$\,0.10$_{-0.64}^{\, +0.62}$ & 1.96\,$\pm$\,0.08$_{-0.28}^{\, +0.40}$ & 2.86\,$\pm$\,0.08$_{-0.26}^{\, +0.34}$ & 2.56\,$\pm$\,0.05$_{-0.68}^{\, +0.30}$  & 2.89\,$\pm$\,0.08$_{-0.70}^{\, +0.45}$ & 2.78\,$\pm$\,0.06$_{-0.68}^{\, +0.68}$ \\[1mm]
  $\log_{10}(M_\ast)$ [$\log_{10}(\rm M_\odot)$] & 10.93\,$\pm$\,0.04$_{-0.47}^{\, +0.35}$ & 10.99\,$\pm$\,0.05$_{-0.50}^{\, +0.30}$ & 10.74\,$\pm$\,0.10$_{-0.50}^{\, +0.31}$ & 11.06\,$\pm$\,0.03$_{-0.68}^{\, +0.30}$  & 11.18\,$\pm$\,0.03$_{-0.50}^{\, +0.35}$ & 11.09\,$\pm$\,0.04$_{-0.54}^{\, +0.37}$ \\[1mm]
  $\log_{10}(L_{\rm IR})$  [$\log_{10}(\rm L_\odot)$] &    12.16\,$\pm$\,0.04$_{-0.24}^{\, +0.35}$  & 12.22\,$\pm$\,0.06$_{-0.34}^{\, +0.30}$  & 12.18\,$\pm$\,0.06$_{-0.25}^{\, +0.26}$  & 12.38\,$\pm$\,0.02$_{-0.23}^{\, +0.25}$  & 12.61\,$\pm$\,0.01$_{-0.24}^{\,+0.21}$ & 12.58\,$\pm$\,0.02$_{-0.27}^{\,+0.23}$ \\[1mm]
  $\log_{10}(\rm SFR)$ [$\log_{10}(\rm M_\odot\,yr^{-1})$] & 2.07\,$\pm$\,0.04$_{-0.34}^{\,+0.42}$  & 2.10\,$\pm$\,0.07$_{-0.43}^{\,+0.43}$  & 2.12\,$\pm$\,0.04$_{-0.27}^{\,+0.39}$  &2.30\,$\pm$\,0.02$_{-0.25}^{\,+0.27}$  & 2.54\,$\pm$\,0.01$_{-0.26}^{\,+0.27}$ & 2.54\,$\pm$\,0.01$_{-0.29}^{\,+0.27}$ \\[1mm]
  $\log_{10}(M_{\rm d})$ [$\log_{10}(\rm M_\odot)$] & 8.23\,$\pm$\,0.02$_{-0.22}^{\,+0.16}$  & 8.25\,$\pm$\,0.03$_{-0.25}^{\,+0.16}$  &  8.16\,$\pm$\,0.04$_{-0.16}^{\,+0.07}$  & 8.74\,$\pm$\,0.02$_{-0.20}^{\,+0.18}$  & 9.04\,$\pm$\,0.01$_{-0.17}^{\,+0.21}$ & 9.04\,$\pm$\,0.01$_{-0.18}^{\,+0.20}$  \\[1mm]
  $A_V $ & 2.42\,$\pm$\,0.13$_{-0.85}^{\,+1.38}$  &  2.77\,$\pm$\,0.08$_{-0.90}^{\,+1.35}$  & 2.43\,$\pm$\,0.24$_{-0.85}^{\,+1.60}$ & 2.32\,$\pm$\,0.23$_{-1.05}^{\,+0.98}$ & 3.06\,$\pm$\,0.07$_{-0.98}^{\,+1.10}$ & 2.85\,$\pm$\,0.06$_{-0.93}^{\,+0.70}$ \\[1mm]
  $\log_{10}(M_{\rm gas}/M_\ast)$ &   $-$0.68\,$\pm$\,0.08$_{-0.55}^{\,+0.54}$ &  $-$0.76\,$\pm$\,0.09$_{-0.44}^{\,+0.64}$ & $-$0.42\,$\pm$\,0.11$_{-0.74}^{\,+0.48}$ & $-$0.26\,$\pm$\,0.04$_{-0.49}^{\,+0.79}$ & $-$0.13\,$\pm$\,0.05$_{-0.44}^{\,+0.66}$ & $-$0.02\,$\pm$\,0.06$_{-0.49}^{\,+0.73}$ \\[1mm]
  $\log_{10}(M_{\rm gas}/{\rm SFR})$ [$\log_{10}({\rm Myrs})$] &   2.17\,$\pm$\,0.04$_{-0.53}^{\,+0.42}$ &  2.10\,$\pm$\,0.10$_{-0.45}^{\,+0.48}$ & 2.18\,$\pm$\,0.09$_{-0.53}^{\,+0.37}$ & 2.49\,$\pm$\,0.02$_{-0.39}^{\,+0.39}$ & 2.55\,$\pm$\,0.02$_{-0.31}^{\,+0.30}$ & 2.55\,$\pm$\,0.03$_{-0.36}^{\,+0.31}$ \\[1mm]
  $\log_{10}(\Delta\rm sSFR)$ &   0.35\,$\pm$\,0.09$_{-0.46}^{\,+0.77}$ &  0.41\,$\pm$\,0.13$_{-0.57}^{\,+0.68}$ & 0.21\,$\pm$\,0.12$_{-0.19}^{\,+0.73}$ & 0.30\,$\pm$\,0.04$_{-0.39}^{\,+0.76}$ & 0.19\,$\pm$\,0.03$_{-0.31}^{\,+0.77}$ & 0.38\,$\pm$\,0.05$_{-0.47}^{\,+0.69}$ \\[1mm]
  $\log_{10}({\rm Age})$ [$\log_{10}(\rm Myrs)$] & 2.84\,$\pm$\,0.06$_{-0.56}^{\,+0.28}$  & 2.97\,$\pm$\,0.05$_{-0.65}^{\,+0.19}$  &  2.63\,$\pm$\,0.06$_{-0.67}^{\,+0.20}$  & 2.67\,$\pm$\,0.03$_{-0.72}^{\,+0.33}$  & 2.58\,$\pm$\,0.03$_{-0.59}^{\,+0.40}$ & 2.54\,$\pm$\,0.05$_{-0.63}^{\,+0.53}$  \\[1mm]
  $\log_{10}(M_{\rm bar})$ [$\log_{10}(\rm M_\odot)$] & $11.0\pm 0.1$  &  $11.1\pm 0.1$  &  $10.8\pm 0.1$  & $11.3\pm 0.1$  & $11.5\pm 0.1$ & ... \\[1mm]
   $\log_{10} (M_{\rm halo})$ [$\log_{10}(\rm M_\odot)$] & $12.5\pm 0.5$ & $12.5\pm 0.5$ & $12.4\pm 0.5$ &  $12.6\pm 0.5$  & $12.8\pm 0.4$  & ... \\[1mm]
  \hline \noalign {\smallskip}
\end{tabular}
\end{table*}

We show the distribution of the $(H-K)$ colour and photometric redshift ($z_{\rm med}$) for the positive and negative sources in the main panel of  Figure~2.  This combination of observed and derived parameters  provides a simple selection of the reddest and most distant counterparts, which includes the majority of real submillimetre galaxies.  To tune the selection we used the properties of the faintest submillimetre sources in the AS2UDS catalogue from \citet{Dudzeviciute20}, with 870-$\mu$m flux densities that overlapped with that of the sources of interest here:  $S_{\rm 870\mu m}$\,$\leq$\,2\,mJy and with $K$-band counterparts.  These are shown in Figure~2a and 2b including their cumulative distributions.  Using the observed distribution of positive and negative source properties and the AS2UDS sample we selected simple cuts on colour and redshift that maximised the total number and completeness of likely true submillimetre sources, while minimising the contamination by likely spuriously-matched field galaxies.

We adopted a cut on the observed colours of  $(H-K)$\,$\geq $\,0.4, as  the cumulative number of positive $K$-band counterparts with red colours rose more rapidly beyond this threshold than for the negative counterparts (see Figure~2a).    We also applied a cut on the photometric redshift of $z_{\rm med}$\,$\geq $\,1 as the cumulative redshift distribution of positive sources similarly  rapidly exceeds the negative sources beyond this redshift, Figure~2b, and only one of 85 AS2UDS sources with $S_{\rm 870\mu m}$\,$\leq$\,2\,mJy had $z_{\rm med}$\,$<$\,1.0.    To estimate the completeness of this  selection we applied the colour and redshift cuts to the AS2UDS sample and determined that  74 of the 84 AS2UDS sources with $K$-band counterparts and $S_{\rm 870\mu m}$\,$\leq$\,2\,mJy had $(H-K)$\,$\geq $\,0.4 and $z_{\rm med}$\,$\geq$\,1.0, corresponding to 88 per cent completeness.

Applying the colour and redshift selection limits to our samples left us with 84 positive and six negative sources, after removing two negative sources with dust masses consistent with zero.   These samples are plotted in Figure~1 illustrating the significant reduction in the expected contamination of the positive sample, as judged by the number of selected negative sources,  from the application of the $(H-K)$ colour and redshift cuts. The number of negative sources indicated a likely contamination rate in the positive sample of around seven per cent, which we viewed as low enough that it would not significantly bias the analysis of their properties.  We discuss below the likely influence on the bulk properties of the sample of both the remaining contamination in the sample and also the incompleteness arising from the selection criteria that were  applied, concluding that these are both less than the statistical uncertainties.

We list the coordinates and properties of the 84 positive sources in the  AS2UDSx sample in Table~1.  We give their unique identification, based on the corresponding AS2UDS ALMA map, 870-$\mu$m coordinates, flux density ($S_{\rm 870\mu m}$) and signal-to-noise of the source,  the coordinates of their $K$-band counterpart, and  the photometric redshift ($z_{\rm med}$), stellar mass ($M_\ast$), star-formation rate (SFR), dust mass ($M_{\rm d}$) and visual attenuation ($A_V$) of the source, along with their associated uncertainties, as derived from the {\sc magphys} fits.  We also flag the nearest matches to the remaining six negative sources, based on proximity in the space of ($S_{\rm 870\mu m}$, $z_{\rm med}$, $M_\ast$ and SFR), to allow a simple statistical correction for contamination if needed.    

We  next sought to identify potential AGN within the AS2UDSx sample.  First by cross-correlating with the X-UDS Chandra survey of \citet{Kocevski18}:  39 of the positive sources lay within the footprint of the  X-ray coverage, but there were no coincident X-ray sources.  We then applied the Spitzer IRAC \citep{Fazio04} AGN colour selection of \citet{Donley12} to the 50 positive sources that have detections in all four IRAC channels.  We removed those at $z_{\rm med}$\,$\geq$\,2.5 where this colour selection becomes ineffective for very dusty sources (see e.g., \citealt{Stach19}) and found 11 remaining sources that fall in the potential AGN region. The only one of these sources that is particularly noteworthy is AS2UDSx0472\_09, which is the brightest 870-$\mu$m source in our faint sample with $S_{\rm 870\mu m}$\,$=$\,2.2\,$\pm$\,0.3\,mJy.  However, as a group the potential IRAC-identified AGN are not distinct in their physical properties,  in particular their stellar masses and star-formation rates are fully consistent with the combined faint AS2UDSx+AS2UDS sample (see Table~2) with median values of $\langle S_{\rm 870\mu m}\rangle $\,$=$\,$0.84\pm 0.14^{\, +0.15}_{-0.38}$\,mJy,  $\langle z_{\rm med}\rangle $\,$=$\,$1.99\pm 0.13^{\, +0.26}_{-0.25}$, $\langle \log_{10}(M_\ast) \rangle$\,$=$\,$10.94\pm 0.15^{\, +0.30}_{-0.59}$, $\langle \log_{10}(\rm SFR) \rangle$\,$=$\,$2.09\pm 0.14^{\, +0.10}_{-0.54}$, $\langle A_V \rangle$\,$=$\,$2.3\pm 0.3^{\, +0.9}_{-0.8}$.   As such we  chose to retain them in our analysis, although we  confirmed that none of the major conclusions qualitatively changed if they were  removed.   We flag the potential IRAC-identified AGNs in Table~1.

Finally, to supplement the faint sample in the analysis of their JWST properties in \S3 we also included a sample of $K$-band-selected massive star-forming galaxies which were detected with ALMA  at  870-$\mu$m from \citet{Tadaki20}.  The physical properties of these sources were derived in an identical manner to the faint AS2UDSx and AS2UDS samples using {\sc magphys} and the same UDS photometric catalogue.
We first removed the ten AS2UDS sources that were reobserved by \citet{Tadaki20} and then apply a SNR\,$\geq$\,3 cut to their ALMA 870-$\mu$m detections.  We then  required $z_{\rm med}$\,$\geq$\,1 -- giving 25 sources with $S_{\rm 870\mu m}$\,$=$\,0.5--1.5\,mJy  (with no overlap with our faint sample) and a further 14 with $S_{\rm 870\mu m}$\,$=$\,2--4\,mJy.  The median properties of the $S_{\rm 870\mu m}$\,$=$\,0.5--1.5\,mJy  sources from \citet{Tadaki20} are comparable to our combined faint AS2UDSx+AS2UDS sample (see Table~2) with $\langle S_{\rm 870\mu m}\rangle $\,$=$\,$0.95\pm 0.09^{\, +0.47}_{-0.33}$\,mJy, $\langle z_{\rm med}\rangle $\,$=$\,$1.96\pm 0.14^{\, +0.40}_{-0.20}$, $\langle \log_{10}(M_\ast) \rangle$\,$=$\,$11.03\pm 0.04^{\, +0.21}_{-0.15}$, $\langle \log_{10}(\rm SFR) \rangle$\,$=$\,$1.62\pm 0.09^{\, +0.40}_{-0.44}$, $\langle A_V \rangle$\,$=$\,$1.9\pm 0.4^{\, +1.4}_{-1.4}$, where we list bootstrap errors on the medians and 16--84$^{\rm th}$ percentile ranges.   The  properties of these sources are indistinguishable from our faint AS2UDSx+AS2UDS sample and so we included them in the morphological analysis below.

%
%
\begin{figure*}
    \centerline{        \psfig{file=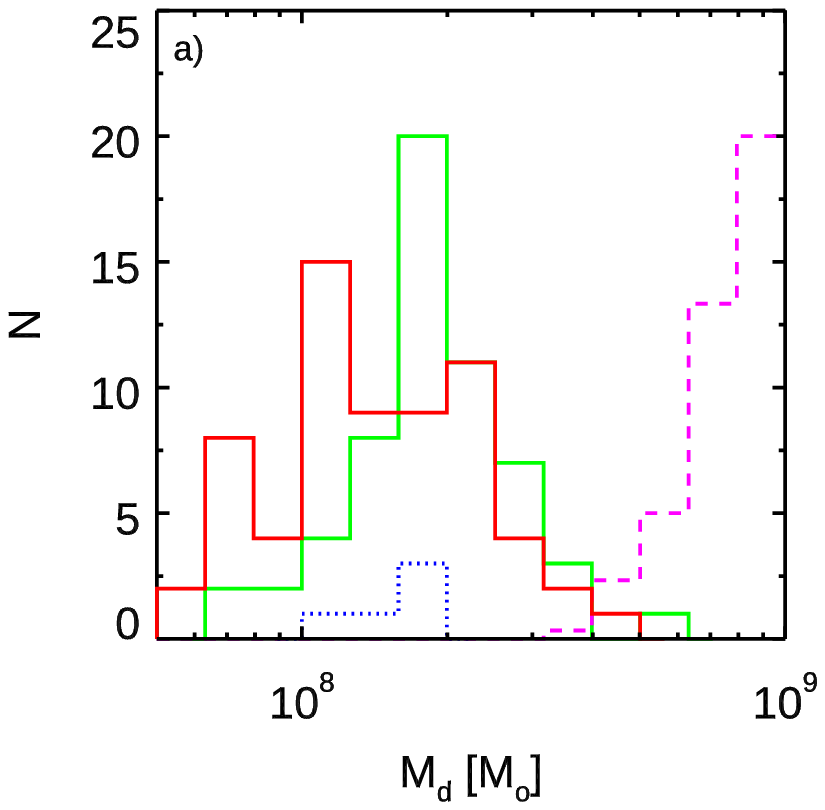, width=2.in, angle=0} \hspace*{-0.37in}
        \psfig{file=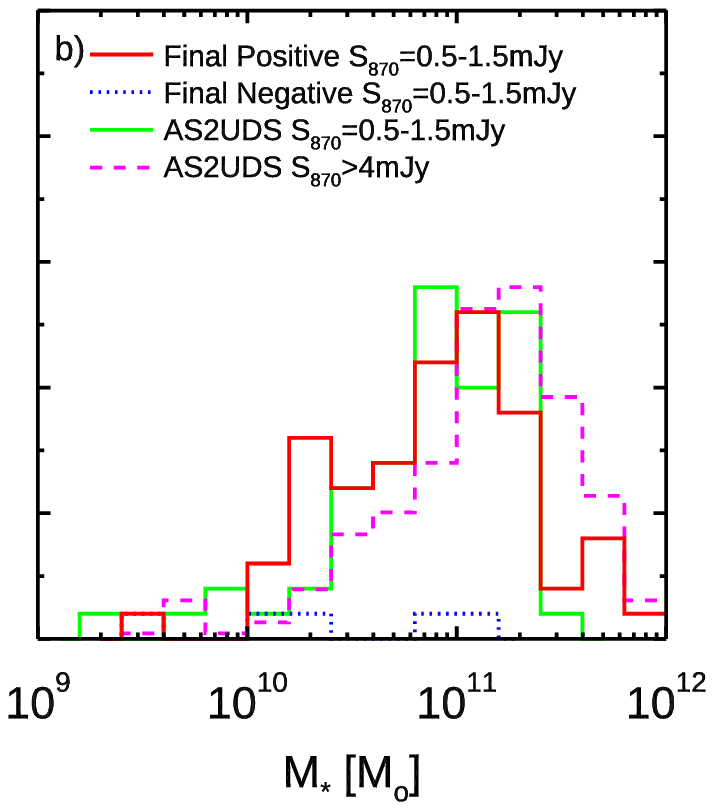, width=2.in, angle=0} \hspace*{-0.37in}
 \psfig{file=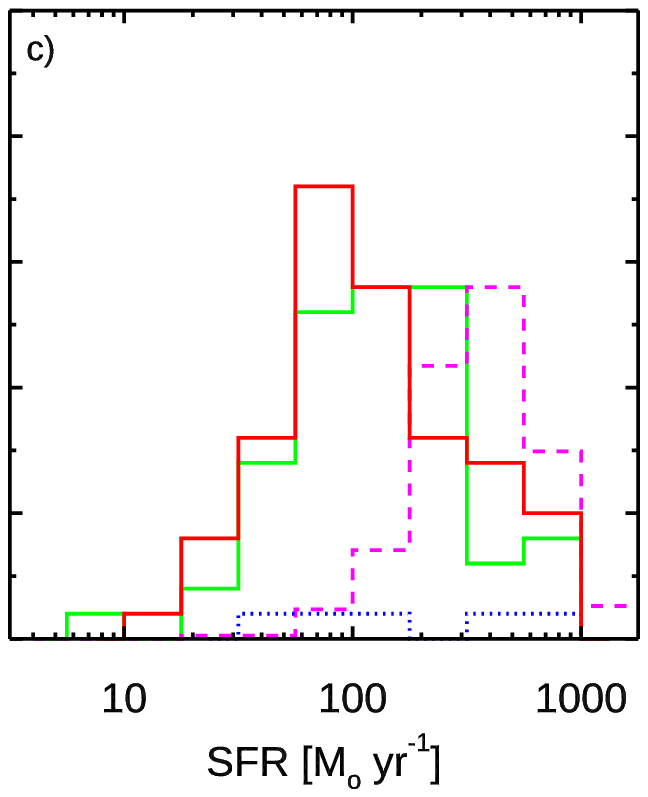, width=2.in, angle=0} \hspace*{-0.37in}
        \psfig{file=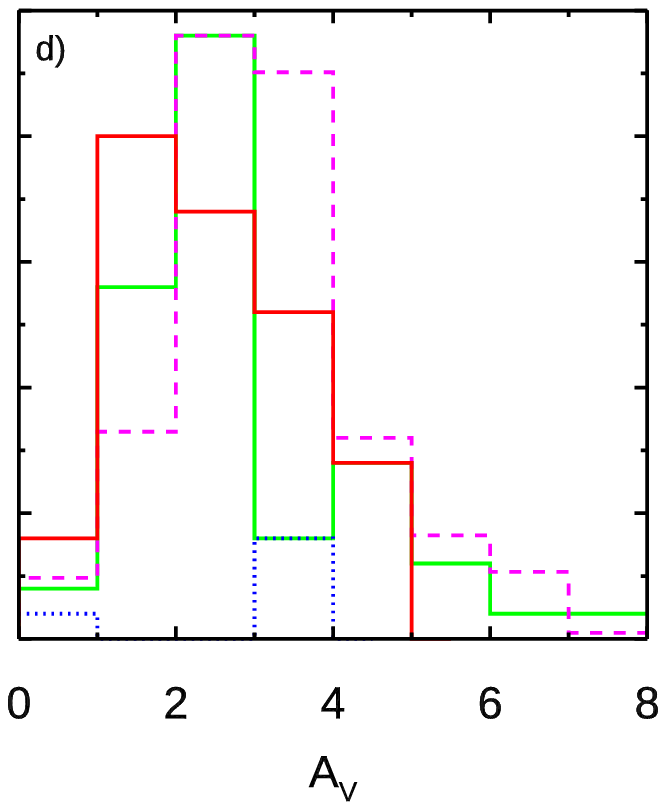, width=2.in, angle=0}} 
      \caption{\small  The distributions of derived physical properties for the final positive and negative AS2UDSx source samples with $S_{\rm 870\mu m}$\,$=$\,0.5--1.5\,mJy:  a) dust mass; b) stellar mass; c) star-formation rate; d) visual attenuation.   These are     compared to  those of the $S_{\rm 870\mu m}$\,$=$\,0.5--1.5\,mJy and $S_{\rm 870\mu m}$\,$\geq$\,4\,mJy  samples from AS2UDS.  The properties of the final positive sample of 65 AS2UDSx sources with $S_{\rm 870\mu m}$\,$=$\,0.5--1.5\,mJy are statistically indistinguishable from the 57 AS2UDS sources at the same flux densities and so they were combined to create a total sample of 122 ALMA-selected sources with $S_{\rm 870\mu m}$\,$=$\,0.5--1.5\,mJy within the UDS field.  These faint submillimetre galaxies have much lower dust masses than the $S_{\rm 870\mu m}$\,$\geq$\,4\,mJy sample from AS2UDS, this reflects the tight relation between 870-$\mu$m flux density and dust mass, as well as lower star-formation rates, but more similar stellar masses and restframe $V$-band attenuations.
}
\end{figure*}

\section{Analysis and Results}

The 716 ALMA maps observed by \citet{Stach19} targeted SCUBA-2 sources in the S2CLS UDS field catalogued by \citet{Geach17}.  The SCUBA-2 catalogue had a false-positive rate of $\sim$\,2 per cent, suggesting that $\sim$\,14 of the 716 maps would not contain any ALMA detected sources. However, \citet{Stach19} actually found 101 maps that lacked an ALMA source above their 4.3-$\sigma$ threshold.

We note that 41 of the 84 positive AS2UDSx sources lie in these 101 ALMA maps which are source-free (or ``blank'') in AS2UDS.    In contrast only one negative source falls in these blank maps.  Hence the  rate of positive AS2UDSx sources  found in previously source-free ALMA maps is 41/84 or 49\,$\pm$\,7 per cent, compared to 1/6 or $\sim$\,17 per cent for the negative sources and an expectation of 101/716 or 14\,$\pm$\,1 per cent for a random distribution.  Thus, the positive AS2UDSx sources are found  $\sim$\,3.5$\times$ more frequently in the  AS2UDS source-free maps than expected from their relative area, while the negative rate is consistent with expected ratio of the AS2UDS source-free and source-containing map areas.   This suggests that many of the faint sources in AS2UDSx are  real. 

Accounting for the  $K$-band coverage of the AS2UDS maps (629 from the 716), the incompleteness in AS2UDSx of order $\sim$\,10 per cent from the $K$-selection, and a $\sim$\,14 per cent correction due to the colour/redshift selection (see \S2),  we estimated that roughly $\sim$\,67 per cent of the AS2UDS source-free maps (58 of 87, after subtracting the expected false-positive SCUBA-2 sources)  contain a $\geq$\,3.1\,$\sigma$ submillimetre source with $S_{\rm 870\mu m}$\,$\sim$\,1\,mJy.  Hence sources similar to those catalogued here would account for the majority of the ``blank'' ALMA maps in \citet{Stach19}.    

Fifteen of the 41 AS2UDSx sources in previously source-free maps lie in just seven ALMA maps which contain more than one positive AS2UDSx source.  Similarly all 43 remaining positive AS2UDSx sources not in previously source-free maps obviously lie in maps with a brighter ALMA companion from the \citet{Stach19} catalogue.  This means that the overall rate of multiplicity in the faint AS2UDSx sample is 69\,$\pm$\,9 per cent, which is significantly higher than the $\sim$\,26 per cent  derived by \citet{Stach18} for the full AS2UDS sample, and even the 44$^{\, +16}_{-14}$ per cent rate found in their brightest $S_{\rm 870\mu m}$\,$\geq$\,9\,mJy sample.   We suggest that the high rate of multiplicity  in AS2UDSx is in large part a result of the fact that we  searched in maps the vast majority, $\sim$\,85 per cent, of which already contained a detected source.  Hence, if the AS2UDSx sources were randomly distributed through the 716 ALMA maps then $\sim$\,85 per cent would lie in a map with another source.    The rate we estimated is lower than this, which can be attributed to the fact that we found that the AS2UDSx sources are more likely to lie in a previously source-free ALMA map.

The combined AS2UDSx+AS2UDS sample includes one map containing four ALMA sources, eight maps containing triples and 125 maps containing doubles.   Accounting for the $K$-band coverage of AS2UDSx, we estimate that 36\,$\pm$\,2 per cent of sources in the combined AS2UDSx+AS2UDS catalogue lie in a field containing another submillimetre galaxy,  with only $\sim$\,7 per cent of the AS2UDS maps remaining as source-free (after accounting for the expected false-positive S2CLS sources).

To understand trends in the submillimetre population with 870-$\mu$m flux density, we first sought to better define our sample
of faint sources and so  applied a flux density cut of   $S_{\rm 870\mu m}$\,$=$\,0.5--1.5\,mJy to isolate a sample of 65 positive and 5 negative sources with median $\langle S_{\rm 870\mu m}\rangle $\,$\sim$\,1\,mJy.   This provides a combined AS2UDSx+AS2UDS sample of 122 ALMA-selected sources with $S_{\rm 870\mu m}$\,$=$\,0.5--1.5\,mJy, effectively doubling the number available the original AS2UDS catalogue of \citet{Stach19}.  The combined faint AS2UDSx+AS2UDS sample is also roughly twice the size of a heterogeneous collection  of  ALMA sources from a  mix of  archival compilations, lensing and blank field surveys at $\sim$\,1\,mm  (43 sources at 870\,$\mu$m  and 29 at 1150\,$\mu$m) available from the literature discussed in the Introduction.   To provide well-defined comparison samples of sources with brighter submillimetre flux densities we also selected sources with $S_{\rm 870\mu m}$\,$=$\,2--4\,mJy and $S_{\rm 870\mu m}$\,$\geq$\,4\,mJy from \citet{Stach19}.

%
%
\begin{figure*}
  \centerline{        \psfig{file=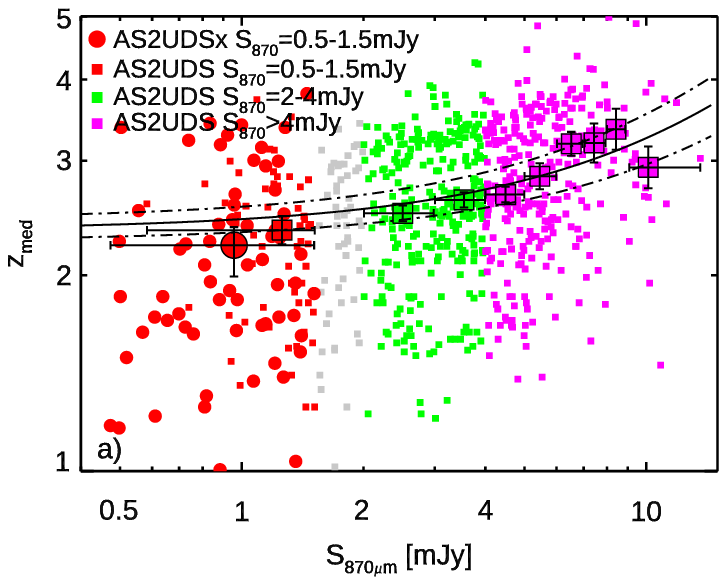, width=3in, angle=0} \psfig{file=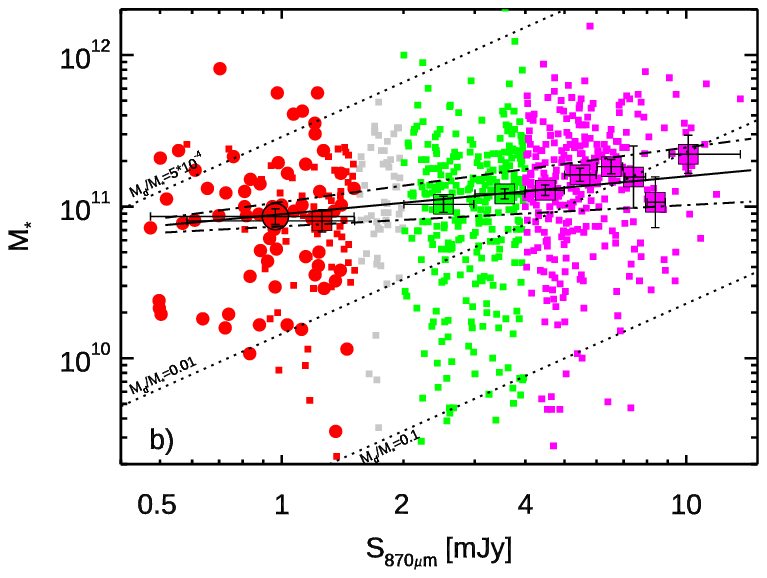, width=3in, angle=0}   }
  \centerline{         \psfig{file=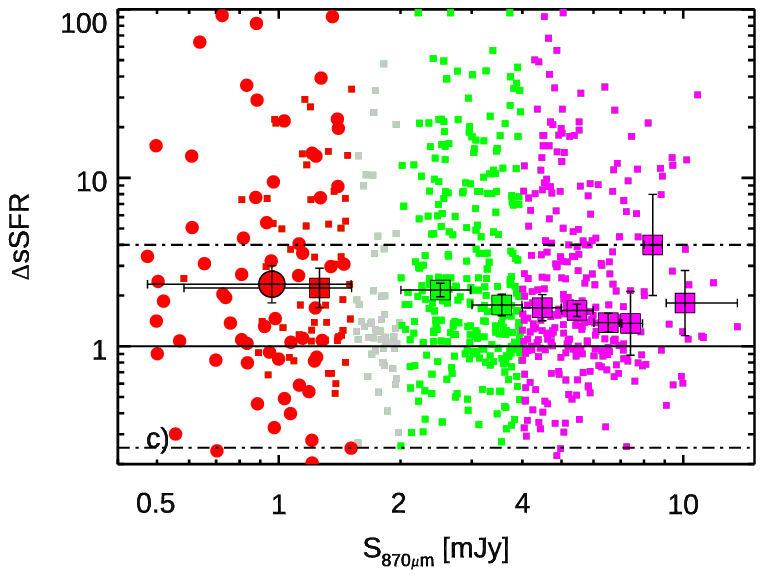, width=3in, angle=0}  \psfig{file=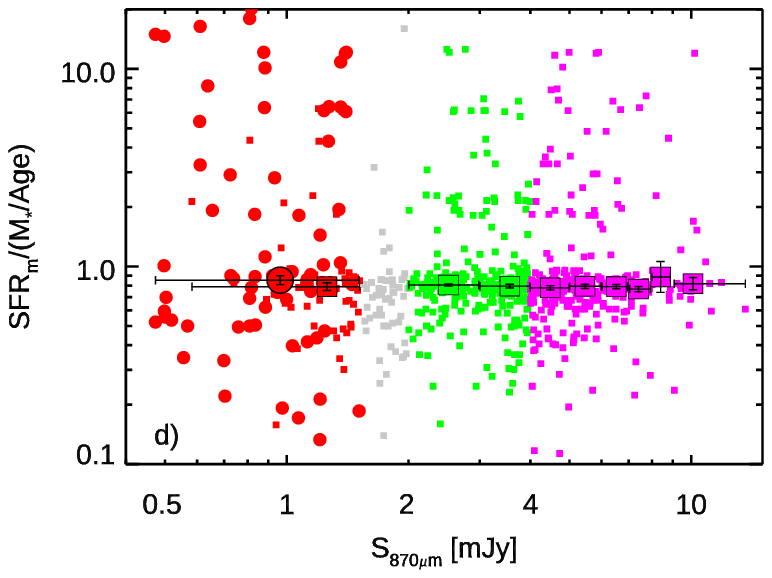, width=3in, angle=0} }
    \centerline{         \psfig{file=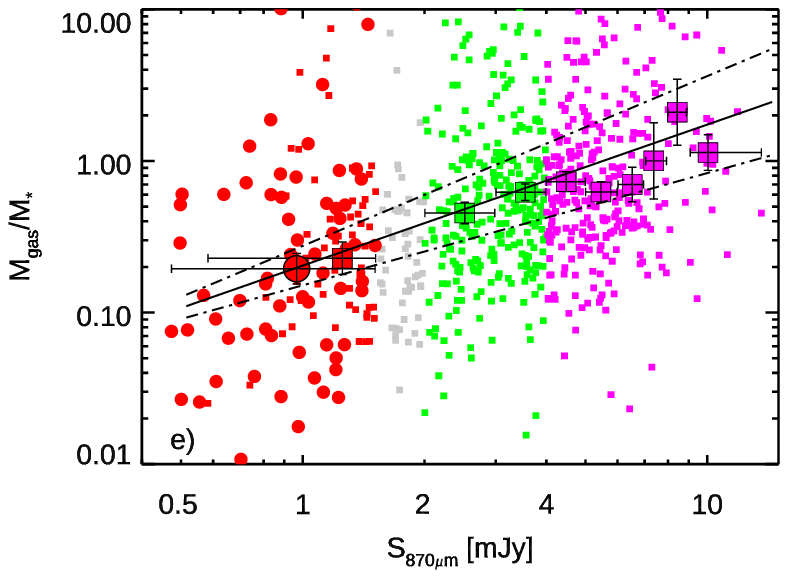, width=3in, angle=0}  \psfig{file=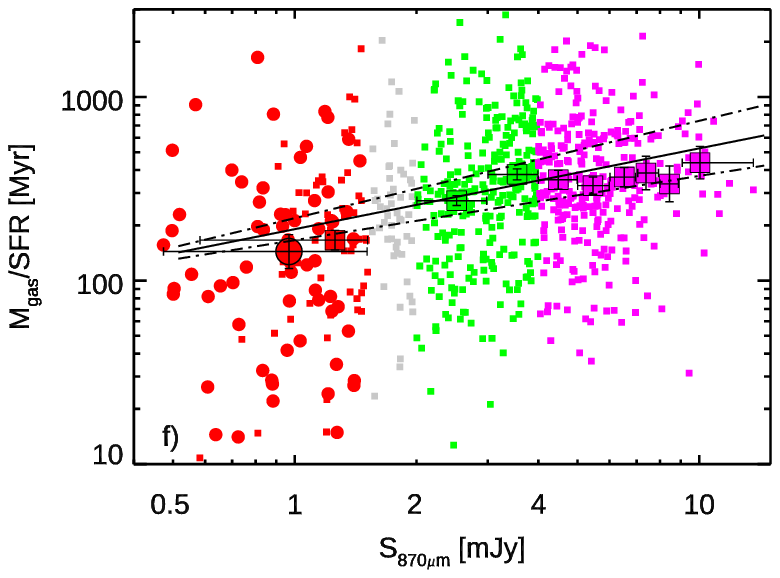, width=3in, angle=0} }
    \caption{\small
      a) The relation between 870-$\mu$m flux density, $S_{\rm 870\mu m}$, and photometric redshift, $z_{\rm med}$ for AS2UDSx and AS2UDS \citep{Stach19, Dudzeviciute20}. Large outlined symbols show medians and bootstrap errors for AS2UDSx and flux-binned AS2UDS. The AS2UDSx median is corrected for $\sim$\,10 per cent incompleteness in high-redshift sources due to the $K$-band requirement (see \S3.1), with the lower redshift error bar indicating this correction. The solid and dot-dashed lines show the best-fit trend and 1-$\sigma$ error from \citet{Stach19}, which the faint AS2UDSx sample broadly follows.
   b) Stellar mass, $M_\ast$, versus $S_{\rm 870\mu m}$ shows a weak trend to lower masses at fainter fluxes: $M_{\ast} \propto S_{\rm 870\mu m}^{0.25}$, a factor $\sim$\,2 across $S_{\rm 870\mu m}$\,$=$\,0.5--10\,mJy. Lines of constant dust-to-stellar mass ratio (0.1, 0.01, $5\times10^{-4}$) indicate slightly lower $M_\ast$ and $M_{\rm d}/M_\ast$ ratios in faint sources.
   c) The offset in sSFR, $\Delta\rm sSFR$, relative to a $K$-selected field sample \citep{Dudzeviciute20}, shows no significant dependence on $S_{\rm 870\mu m}$.
   d) The ratio of current SFR (last $\sim$\,100\,Myr) to time-averaged SFR ($M_\ast/{\rm Age}$) peaks near $\sim$\,1 for bright sources, but fainter sources show $\sim$\,0.4\,dex larger dispersion, suggesting more diverse evolutionary histories.
   e) Gas fraction, $M_{\rm gas}/M_\ast$, declines toward lower $S_{\rm 870\mu m}$, with AS2UDSx and faint AS2UDS samples following the brighter trend.
   f) Depletion time, $M_{\rm gas}/{\rm SFR}$, decreases at fainter flux densities, with the faintest samples broadly consistent with this trend.      }
\end{figure*}

Figure~3 compares the distributions of some of the inferred properties of the  positive and negative AS2UDSx sources with $S_{\rm 870\mu m}$\,$=$\,0.5--1.5\,mJy, along with purely 870-$\mu$m-selected samples in the same and brighter flux density ranges from AS2UDS.  In general  the distributions of properties for the positive AS2UDSx sources are  in good agreement with the similarly faint AS2UDS sources. The most significant mismatch, $2.6\sigma$, is between dust masses: $\langle \log_{10}(M_{\rm d}) \rangle$\,$=$\,$8.12\pm 0.04^{\, +0.23}_{-0.20}$ for the positive AS2UDSx sources compared to $\langle \log_{10}(M_{\rm d}) \rangle$\,$=$\,$8.26\pm 0.02^{\, +0.18}_{-0.15}$ for AS2UDS, due to the higher ALMA significance cut and hence slightly higher median 870-$\mu$m flux density of the latter.  The sample of brighter submillimetre sources from AS2UDS shown in Figure~3 exhibit much higher dust mass and star-formation rates, along with slightly higher stellar masses and visual attenuation, compared to the faint AS2UDSx and AS2UDS samples. We note that  the remaining six faint negative AS2UDSx sources have a spread in properties that confounds attempts to further reduce the residual contamination from false-positive noise sources in the positive sample.   Nevertheless, this broad spread in their properties means that we do not expect the remaining false-positive sources to bias any of the median properties we derive for our final sample and so we do not consider the negative sources further.

We list the median properties (and 16--84$^{\rm th}$ percentile ranges) of the various 870-$\mu$m-selected, flux-limited samples  in Table~2.  These include the number of sources, the median 870-$\mu$m flux densities, photometric redshifts, stellar masses, 8--1000\,$\mu$m luminosities, star-formation rates, dust masses, $V$-band attenuation, gas fraction ($M_{\rm gas}/M_\ast$, where the molecular gas masses use   gas-to-dust ratios following \citealt{GomezGuijarro22b}), gas depletion timescale ($M_{\rm gas}/{\rm SFR}$), specific star-formation rate relative to the ``main sequence'' at their redshift ($\Delta\rm sSFR$), age, baryonic masses (sum of stellar and molecular gas masses) and halo masses (see \S 3.4).

\subsection{Incompleteness and surface density}

We  assessed the  incompleteness in the faint AS2UDSx  galaxy sample due to the requirement for a $K$-band counterpart by using the comparably faint submillimetre galaxies in the AS2UDS catalogue. There are 54 AS2UDS  galaxies with  $S_{\rm 870\mu m}$\,$=$\,0.5--1.5\,mJy in the $K$-band footprint of the UDS, of which five have no $K$-band counterpart brighter than $K$\,$\leq$\,25.7.  This indicates a likely incompleteness of 9\,$\pm$\,4 per cent for the faint AS2UDSx sample.  We note that this rate of faint submillimetre galaxies with $S_{\rm 870\mu m}$\,$\sim$\,1\,mJy and $K$\,$\geq$\,25.7 of $\sim$\,9\,$\pm$\,4  per cent  is similar  to the  $\sim$\,15\,$\pm$\,2 per cent rate estimated for $S_{\rm 870\mu m}$\,$\geq$\,3.6\,mJy submillimetre galaxies by \citet{Smail21}, suggesting no strong variation with 870-$\mu$m flux density.  This likely reflects two opposing effects on the $K$-band brightness of counterparts.  Firstly, a lower median redshift for fainter submillimetre sources (see Figure~4a), but also somewhat lower stellar masses (Figure~4b) and lower dust attenuation (Table~2).  When comparing to the brighter, purely 870-$\mu$m-selected,  samples in our analysis below we  crudely assess the influence of this incompleteness  by applying a $K$-band selection to the brighter submillimetre samples (see Table~2).   Overall, we expect that the requirement for a $K$-band counterpart in AS2UDSx  biases this sample towards somewhat lower redshifts, higher stellar masses and lower $A_V$, than a purely 870-$\mu$m-selected sample, although the inclusion  of the $(H-K)$\,$\geq $\,0.4 cut in the selection may counteract the bias in $A_V$, which is further diluted by combining the AS2UDSx and AS2UDS samples.

Corrected for the  contamination from remaining noise sources in the positive AS2UDSx sample using the number of negative sources, there are 60 positive sources with an incompleteness of $\sim$\,10 per cent due to the $K$-selection and a $\sim$\,14 per cent correction due to the colour/redshift selection (see \S2), suggesting a true population of  $\sim$\,75\,$\pm$\,15.  We estimate that for a typical source with $S_{\rm 870\mu m}$\,$\sim$\,1\,mJy and a SNR$_{0.5}$\,$\sim$\,3.1 we have an effective survey area of 28.5 arcmin$^{2}$, although only 21.2 arcmin$^2$ of this has the necessary $K$-band coverage.   We  add to our AS2UDSx sample the  54 sources with $S_{\rm 870\mu m}=$\,0.5--1.5\,mJy  from AS2UDS in this area giving a total of 129\,$\pm$\,19.  This indicates a surface density of $S_{\rm 870\mu m}$\,$\sim$\,1\,mJy sources of (2.2\,$\pm$\,0.3)\,$\times 10^4$ degrees$^{-2}$, which falls in the range of values of (2\,$\pm$\,1)\,$\times$\,10$^4$ degrees$^{-2}$ estimated from narrow blank-field or cluster lensing surveys with ALMA at 1.2-mm by
\citet{Hatsukade18}, \citet{GonzalezLopez20} and \citet{Fujimoto24}\footnote{To convert wavebands for the number counts we use the relation $S_{\rm 870\mu m}/S_{\rm 1150\mu m}=2.3-0.04 \times S_{\rm 870\mu m}$ based on the sample \citet{Dudzeviciute20} and \citet{Ikarashi17} which gives a flux correction factor of 0.44.}.

We had  expected that the source density in our sample would be  higher than that in blank field surveys due to the multiplicity bias arising  from blending in  low-resolution, single-dish surveys \citep[e.g.,][]{Karim13,Stach18}.  However, the good agreement in the counts indicates that by $S_{\rm 870\mu m}$\,$\sim$\,1\,mJy this bias may be declining as the surface densities are   approaching one per 14.5$''$ FWHM SCUBA-2 beam (equivalent to $\sim$\,8\,$\times 10^4$ degrees$^{-2}$).   We should also stress that the faint sample is severely under-represented compared to a true blank field survey:  while the combined AS2UDS+AS2UDSx sample contains roughly the expected  number of $S_{\rm 870\mu m}$\,$\geq$\,4\,mJy sources for the area of the S2CLS UDS field \citep{Geach17},  $\sim$\,0.96 degrees$^{-2}$  analysed by \citet{Stach19}, and around $\sim$\,60 per cent of the $S_{\rm 870\mu m}$\,$=$\,2--4\,mJy sources (due to incompleteness), our sample  contains just $\sim$\,0.6 per cent of the $S_{\rm 870\mu m}$\,$\sim$\,1\,mJy sources in this region.

%
%
\begin{figure*}
  \centerline{        \psfig{file=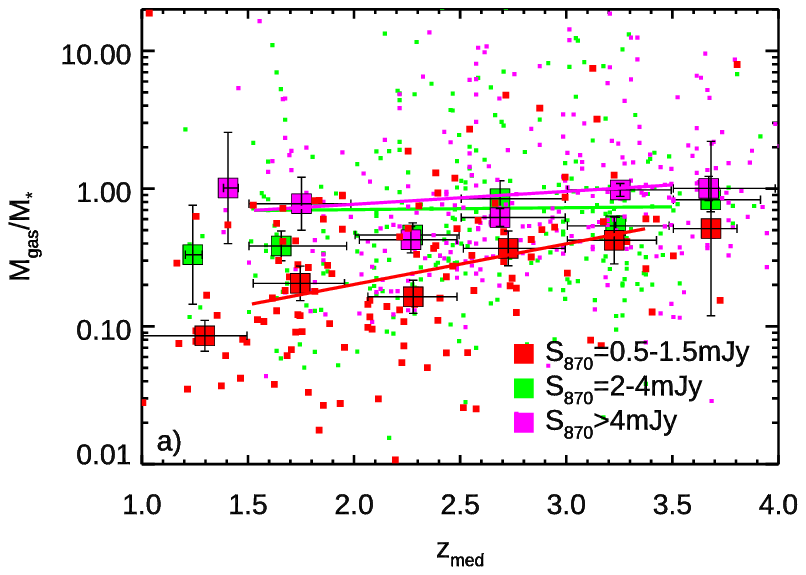, width=2.4in, angle=0} \hspace*{-0.35cm}\psfig{file=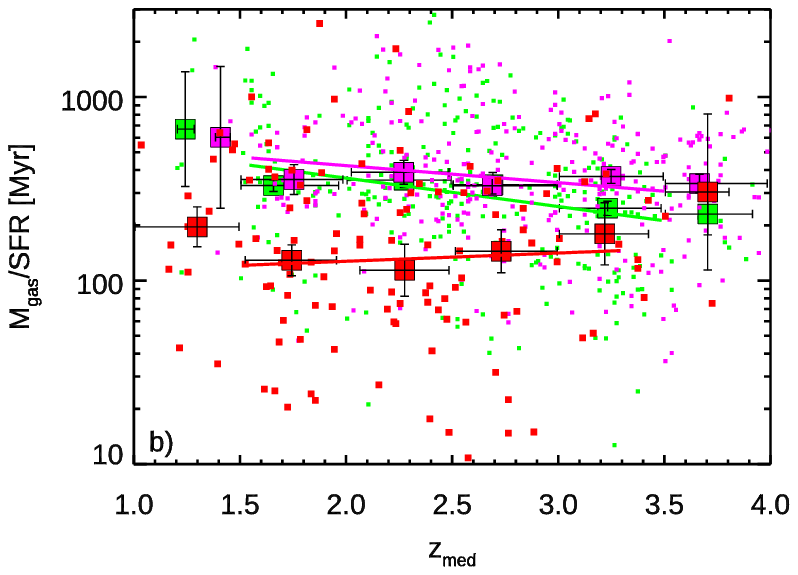, width=2.4in, angle=0}
 \hspace*{-0.45cm} \psfig{file=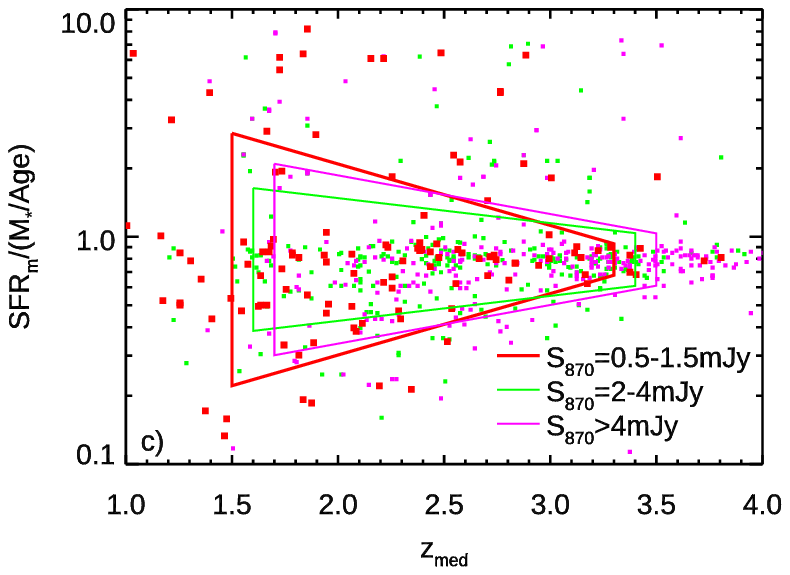, width=2.4in, angle=0} }
    \caption{\small
      a) Gas fraction versus redshift for the $S_{\rm 870\mu m}$\,$=$\,0.5--1.5\,mJy (using the combined AS2UDSx and AS2UDS samples), $S_{\rm 870\mu m}$\,$=$\,2-4\,mJy  and $S_{\rm 870\mu m}$\,$\geq$\,4\,mJy subsets.      Large symbols show the median and bootstrap error bars for  $\Delta z_{\rm med}$\,$=$\,0.5 bins. The lines are robust linear fits to the running medians of the corresponding sample using a five-source window in the range $z_{\rm med}$\,$=$\,1.5--3.5 which is well populated by all three samples.
      b) Depletion timescale versus redshift for the $S_{\rm 870\mu m}$\,$=$\,0.5--1.5\,mJy, $S_{\rm 870\mu m}$\,$=$\,2--4\,mJy  and $S_{\rm 870\mu m}$\,$\geq$\,4\,mJy subsets.  
      c) Ratio of the current SFR to the average SFR over the lifetime of the galaxy for the three samples with the dispersion indicated by polygons encompassing the absolute deviations of the samples in the redshift ranges $z_{\rm med}$\,$=$\,1.5--2.5 and $z_{\rm med}$\,$=$\,2.5--3.5.  All three flux ranges show a broader range in star-formation histories at lower redshifts, but the faintest sources have the largest dispersion in star-formation histories, hinting at a wider variety of evolutionary paths.
}
\end{figure*}

\subsection{Trends in properties with 870-$\mu$m flux density and redshift}

We next turned to our goal of searching for variations in the properties of submillimetre galaxies and less active populations as a function of submillimetre flux density.  We first compared the tabulated properties in Table~2 for the $S_{\rm 870\mu m}\sim$\,1\,mJy and $S_{\rm 870\mu m}\geq$\,4\,mJy samples finding that the most significant differences are that the faint submillimetre galaxies show  lower median redshifts, star-formation rates (or equivalently $L_{\rm IR}$) and $A_V$,  along with marginally lower stellar masses (as well as the expected lower dust-masses directly due to their selection on $S_{\rm 870\mu m}$).   In general the properties of the  $S_{\rm 870\mu m}$\,$=$\,2--4\,mJy sources fall between the two extremes.

To assess the potential influence of the $K$-band selection in AS2UDSx we applied a similar requirement for a $K$-band counterpart on the 
$S_{\rm 870\mu m}$\,$\geq$\,4\,mJy sample in the final column of Table~2.   This had a relatively small influence on all of the properties apart from $A_V$ and $\Delta\rm sSFR$.  Although we note that the reliability of this test may be weakened if, as expected, fainter submillimetre galaxies have lower stellar masses and hence potentially fainter $K$-band counterparts at high redshifts than the brighter subset.   However, we also note that the bias arising from the requirement of a $K$-band counterpart in AS2UDSx is diluted in our $S_{\rm 870\mu m}$\,$=$\,0.5--1.5\,mJy sample as around half of the sources are from the purely 870-$\mu$m-selected AS2UDS sample\footnote{We do not expect the redshift and $(H-K)$ colour selections to strongly bias the median redshifts or star-formation rates of the faint sources, and hence their agreement with the trends seen in the wider population should be robust.   However, there is a bias towards higher values of dust attenuation  due to the red $(H-K)$ colour cut.  For the AS2UDS sources in Figure~2c we find  $\langle A_V\rangle$\,$=$\,1.7\,$\pm$\,0.9 for the $\sim$\,13 per cent with   $(H-K)$\,$<$\,0.4, compared to $\langle A_V\rangle$\,$=$\,2.6\,$\pm$\,1.0 for the remaining $\sim$\,87 per cent  of redder selected sources, with the whole sample having $\langle A_V\rangle$\,$=$\,2.7\,$\pm$\,0.9.  However, this minor  bias is also in the opposite sense of the expected bias due to  the need for a $K$-band counterpart (see Table~2), mitigating its effect somewhat.  There may also be a weak bias against lower-mass sources due to the colour cut, but again the requirement for a $K$-band counterpart for the AS2UDSx sources acts in the opposite sense (see Table~2). }.

To quantify the  differences in star-formation rate and dust mass in the different   870-$\mu$m flux density samples,  we  exploited the empirical relationship derived between these parameters from fitting a plane to the $S_{\rm 870\mu m}$--$M_{\rm d}$--$L_{\rm IR}$ distribution of bright submillimetre galaxies by Ikarashi et al.\ (in prep.), following  \citet{Liang19}:
$$ \log_{10}(S_{\rm 870\mu m}/{\rm mJy}) \propto 0.28 \times \log_{10} (L_{\rm IR}/{\rm L_\odot}) + 0.68 \times \log_{10} (M_{\rm d}/{\rm M_\odot})  $$

Comparing the combined AS2UDSx+AS2UDS $S_{\rm 870\mu m}$\,$=$\,0.5--1.5\,mJy sample with the $K$-detected $S_{\rm 870\mu m}$\,$\geq$\,4\,mJy sample from AS2UDS and using the relation above the ratios of  $\log_{10}(L_{\rm IR})$\footnote{We use $\log_{10}(L_{\rm IR})$ from {\sc magphys}, but note that the scaling from SFR listed in Table~2 is $\log_{10}(L_{\rm IR/SFR})$\,$=$\,10.05\,$\pm$\,0.02$^{\, +0.19}_{-0.13}$.}  and $\log_{10}(M_{\rm d})$ for the two samples implies factors of  2.6$^{0.28}$ \,$\sim$\,1.3 and 6.5$^{0.68}$\,$\sim$\,3.6,
from the differences in $L_{\rm IR}$ and $M_{\rm d}$ respectively.  These factors when combined   predict a median ratio of $S_{\rm 870\mu m}$ between the bright and faint samples of 4.7, which is in very good agreement with the observed ratio of 4.6\,$\pm$\,0.2.  Thus the combination of the estimated  far-infrared luminosities (or SFRs) and dust masses of the faint and bright submillimetre galaxies is consistent with an empirical model where the variation in their dust masses  is responsible for $\sim$\,80 per cent of their relative 870-$\mu$m flux densities.    

To visually assess some of the main trends with 870-$\mu$m flux density (reflecting a combination of their dust masses and far-infrared luminosities) in the submillimetre population, in Figure~4 we compared 
the  properties of the faint  AS2UDSx sources to those from AS2UDS with $S_{\rm 870\mu m}$\,$=$\,0.5--1.5\,mJy, $S_{\rm 870\mu m}$\,$=$\,2--4\,mJy  and $S_{\rm 870\mu m}$\,$\geq$\,4\,mJy.  Figure~4a shows the variation in photometric redshift with 870-$\mu$m flux density, with the average redshift of the three flux-limited samples at $z_{\rm med}$\,$\sim$\,2.5, and fainter submillimetre sources typically lying at lower redshifts.  The faint AS2UDSx (and AS2UDS) samples broadly follow the empirical relation suggested by \citet{Stach19}: $z_{\rm med}=(2.35\pm 0.09) + (0.09\pm 0.02) S_{\rm 870\mu m}$, based on a fit to the AS2UDS sample. While for the  inverse trend, of $S_{\rm 870\mu m}$ with redshift, the binned medians were fitted by $S_{\rm 870\mu m} = (0.40\pm 0.12) \times  (1+z_{\rm med})^{1.7\pm 0.2}$ across $z$\,$=$\,1--4.

Figure~4b shows the variation in stellar mass, $M_\ast$, with $S_{\rm 870\mu m}$.  There is a weak gradient ($\sim$\,2.2\,$\sigma$) in the sense of lower stellar masses at lower 870-$\mu$m flux densities: $M_{\ast} \propto S_{\rm 870\mu m}^{0.25\pm 0.11}$.  This amounts to a factor of $\sim$\,2 decline in stellar mass from $S_{\rm 870\mu m}$\,$\sim$\,10\,mJy to $S_{\rm 870\mu m}$\,$\sim$\,0.5\,mJy.
Plotted on Figure~4b are lines of constant dust-to-stellar mass ratio. These indicate that the faint AS2UDSx+AS2UDS $S_{\rm 870\mu m}$\,$=$\,0.5--1.5\,mJy galaxies  have lower average dust-to-stellar ratios than the brighter populations, and potentially therefore lower gas fractions.

%
%
\begin{figure*}
\centerline{
\psfig{file=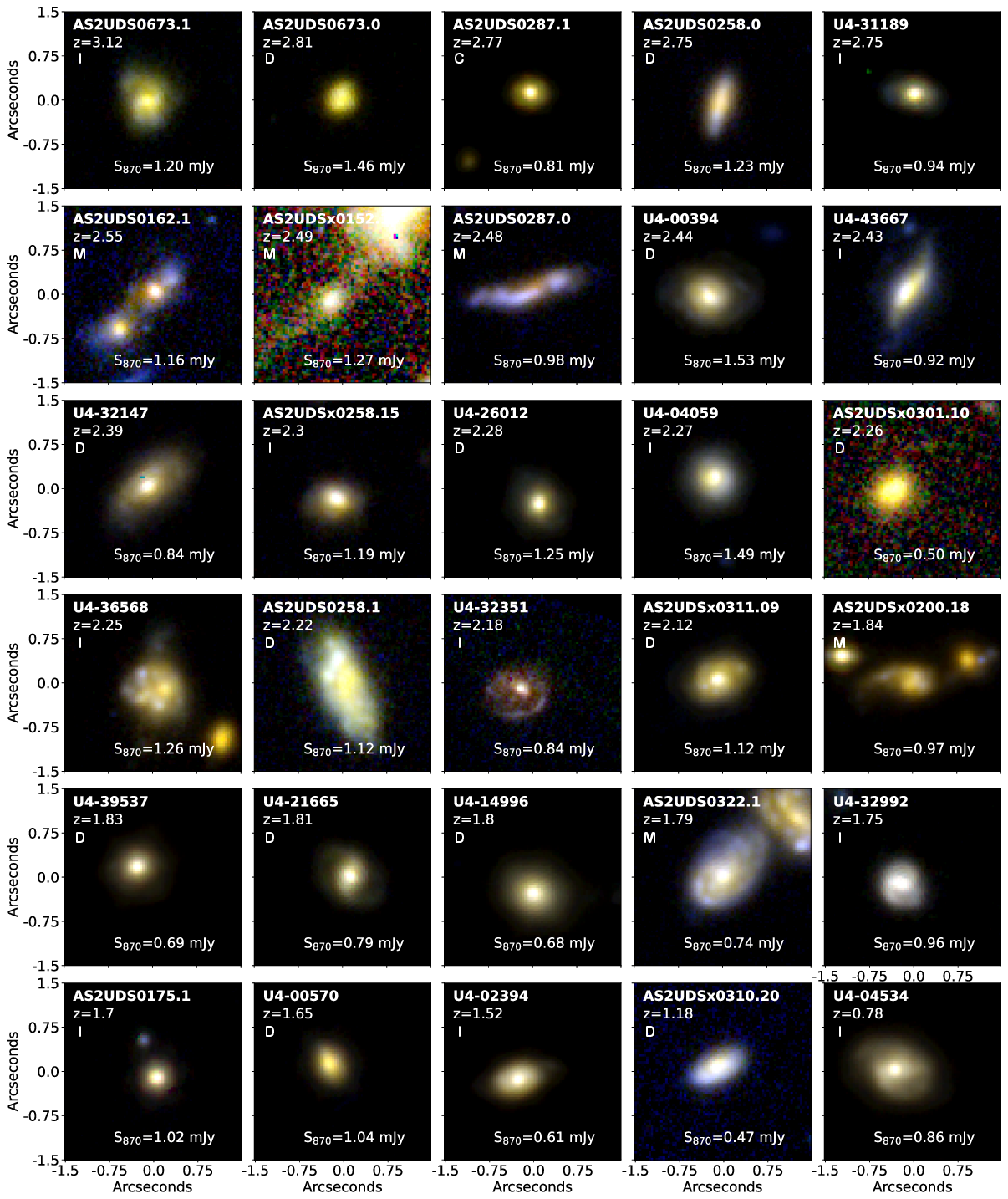,width=6.in, angle=0}}
~\vspace*{-0.8in}\caption{\small  JWST NIRCam imaging of $S_{\rm 870\mu m}$\,$=$\,0.5--1.5\,mJy sources in the UDS field ranked by redshift.  These colour images use F200W for the blue channel, F356W for  green  and F444W for the red (with the exception of U4-32992 which uses F277W for blue and U4-32351 with uses F115W/F150W/F200W for blue, green and red respectively).  These include the AS2UDSx sample identified in this work, supplemented by AS2UDS sources  \citep{Stach19} and similar sources selected from \citet{Tadaki20}, identified by ``U4'' prefixes, see \S2.  Each thumbnail is 3$''$\,$\times$\,3$''$ or $\sim$\,25\,kpc square at their typical redshifts and is centered on the submillimetre position to illustrate the uncertainty in these for the lower SNR sources. We label them with their photometric redshift and 870-$\mu$m flux densities, as well as their visually classifications as major mergers (M), minor mergers or interactions (I), regular discs (D), or compact (C) following \citet{Gillman24}.   These images use the same filters and scaling as those of the  $S_{\rm 870\mu m}$--brighter examples shown in Figure~1 of \citet{Gillman24} which they can be compared to.  We find that the distribution of visual morphological classes  is statistically indistinguishable between our $S_{\rm 870\mu m}$\,$=$\,0.5--1.5\,mJy sample and the $S_{\rm 870\mu m}$\,$\geq$\,2\,mJy sources in \citet{Gillman24}.  This shows that there is no strong variation in the rate of major or minor mergers/interactions in the  galaxy population with submillimetre flux density.   
}
\end{figure*}

In Figure~4c and 4d we show the distributions of $\Delta \rm sSFR$ and SFR/($M_\ast/\rm Age$) with 870-$\mu$m flux density.   $\Delta  \rm sSFR$ is calculated from the ratio of the specific star-formation rate of each source to that of the fit to the trend in sSFR with redshift in the $K$-band selected massive, star-forming field sample in the UDS with $M_\ast$\,$\gs$\,3.5\,$\times$\,10$^{10}$\,M$_\odot$  from \citet{Dudzeviciute20}.  The trend in median sSFR with redshift in this field sample is consistent with that reported by \citet{Karim11} of sSFR\,$\propto (1+z)^{3.6\pm 0.2}$.      SFR/($M_\ast/\rm Age$) is a measure of the current SFR to the average over the lifetime of the galaxy, high ratios indicate that the galaxy is currently undergoing a significant burst of star formation, while low ratios suggest the star-formation rate is declining to the observed epoch.

None of the submillimetre galaxy samples show any statistically-significant variation in either $\Delta \rm sSFR$ or SFR/($M_\ast/\rm Age$) with redshift. The median $\Delta \rm sSFR$ is $\langle \Delta \rm sSFR\rangle$\,$=$\,1.7\,$\pm$\,0.2, indicating that on-average submillimetre galaxies lie above the so-called ``main sequence'' with $\sim$\,30 per cent of them falling in the  ``starburst'' regime, $\Delta \rm sSFR$\,$\geq$\,4, at all submillimetre flux densities.  The ratio of current star-formation rate (over the last $\sim$\,100\,Myrs)  to the time-average rate for each galaxy ($M_\ast/{\rm Age}$) is also  constant with $S_{\rm 870\mu m}$, showing  a strong peak around $\sim$\,1 for bright sources (indicating that the current star-formation events have  contributed significantly to the galaxies's masses), but with a $\sim$\,0.2-dex broader dispersion in the fainter $S_{\rm 870\mu m}$\,$=$\,0.5--1.5\,mJy sample (both flux ranges also have similar median ages, see Table~2).

The final two panels of Figure~4 illustrate the variation in gas fraction, $M_{\rm gas}/M_\ast$ (Figure~4e), and depletion time scale,  $M_{\rm gas}/\rm SFR$ (Figure~4f), with 870-$\mu$m flux density.  Both show relatively strong trends with typically lower gas fractions and shorter depletion timescales seen in fainter submillimetre samples (with the AS2UDSx and AS2UDS $S_{\rm 870\mu m}$\,$=$\,0.5--1.5\,mJy samples in agreement with these trends).   The power-law relations fitted to the binned median values are:  $\log_{10}(M_{\rm gas}/M_\ast)$\,$=$\,$-$0.71\,$\pm$\,0.11\,$+$\,(0.87\,$\pm$\,0.16)\,$\log_{10}(S_{\rm 870\mu m})$ and $\log_{10}(M_{\rm gas}/\rm SFR)$\,$=$\,2.24$\pm$\,0.06\,$+$\,(0.41\,$\pm$\,0.09)\,$\log_{10}(S_{\rm 870\mu m})$ for the gas fraction and depletion timescales (in units of Myrs) respectively.     Part of the strength of these trends is likely driven by the tight relation between our estimate of $M_{\rm gas}$, which is based on dust mass from {\sc magphys}, and 870-$\mu$m flux density, which is major driver of $M_{\rm d}$ (see above).  This is illustrated by the relative strength of the trends in Figure~4b compared to Figure~4e.

In summary, Figure~4   shows that the faintest submillimetre galaxies, with $S_{\rm 870\mu m}$\,$\sim$\,1\,mJy, have  properties that broadly follow the trends seen in the brighter population.   They lie at slightly lower redshifts than the brighter submillimetre galaxies and have slightly lower stellar masses, with lower dust-to-stellar ratios (and hence likely lower gas fractions) and gas depletion timescales, but similar current-to-past average star-formation rates.   These properties suggest that the faint population are more ``evolved'' than the brighter sources, having less gas available for future star formation, although there are also hints that they may have more diverse evolutionary histories than the brighter population.  We next look in more detail at how these properties vary as a function of redshift to identify trends  that may reflect different evolutionary behaviour.

The average redshift of the three flux-selected submillimetre samples in Table~2  is $\langle z_{\rm med} \rangle$\,$\sim$\,2.5 and so we divide the samples in  redshift  at this point.  One consequence of the  variation in $S_{\rm 870\mu m}$ with redshift shown in Figure~4a is that   the $S_{\rm 870\mu m}$\,$=$\,0.5--1.5\,mJy sample has few sources above $z_{\rm med}$\,$\sim$\,3.5 and the $S_{\rm 870\mu m}$\,$\geq$\,4\,mJy sample has few below $z_{\rm med}$\,$\sim$\,1.5. Hence we choose to focus our analysis on comparing and contrasting the populations in the redshift ranges:  $z_{\rm med}$\,$=$\,1.5--2.5 and $z_{\rm med}$\,$=$\,2.5--3.5, where all three samples of flux-selected sources have relatively robust statistics.

We show in Figure~5 the variation in gas fraction, depletion timescale and current-to-lifetime-average star-formation rate, as a function of redshift for the  combined AS2UDSx and AS2UDS $S_{\rm 870\mu m}$\,$=$\,0.5--1.5\,mJy sample and the two brighter AS2UDS samples.   In the first two panels, showing $M_{\rm gas}/M_\ast$ and $M_{\rm gas}/{\rm SFR}$,  we plot the median values in $\Delta z_{\rm med}$\,$=$\,0.5 bins for each sample and show linear fits to five-source running medians to illustrate the broad trends  across  $z_{\rm med}$\,$=$\,1.5--3.5.

Splitting the redshift range  in Figure~5a and Figure~5b in at $z_{\rm med}$\,$=$2.5,
we see no statistically significant differences as a function of redshift in the median gas fractions or depletion timescales
of the brighter two samples, $S_{\rm 870\mu m}$\,$=$\,2--4\,mJy  and $S_{\rm 870\mu m}$\,$\geq$\,4\,mJy, with both showing slightly rising gas fractions and slightly declining gas consumption timescales at higher redshifts.    The median properties as a function of redshift  in the combined AS2UDSx and AS2UDS $S_{\rm 870\mu m}$\,$=$\,0.5--1.5\,mJy sample hints at somewhat different behaviour.  At higher redshifts they show less marked differences from the brighter populations, but at lower redshifts, below $z_{\rm med}$\,$\sim$\,2.5, they exhibit increasingly distinct properties with  lower gas fractions and shorter gas consumption timescales.

So, for example, at $z_{\rm med}$\,$=$\,1.5--2.5 there is a  4.1-$\sigma$ difference in gas fraction
between the faint and the brighter samples, compared to 2.0-$\sigma$ at $z_{\rm med}$\,$=$\,2.5--3.5.
We quantify this with a likelihood ratio test using Kolmogorov-Smirnov statistics showing that the $S_{\rm 870\mu m}$\,$=$\,0.5--1.5\,mJy sample at $z_{\rm med}$\,$=$\,1.5--2.5  has gas fractions and depletion timescales that have much lower probabilities of being drawn from the same parent population as the brighter samples at their respective redshifts  of   $L\!R$\,$\sim$\,10$^{-4.2}$ and $\sim$\,10$^{-10.4}$ respectively, compared to the similar comparison of the populations at $z_{\rm med}$\,$=$\,2.5--3.5.

In the final panel of Figure~5 we focus on the diversity in the current-to-lifetime-average star-formation rate as a function of redshift.  We measure the median dispersion in each flux-limited sample at $z_{\rm med}$\,$=$\,1.5--2.5 and $z_{\rm med}$\,$=$\,2.5--3.5 and plot polygons to indicate their variation with median redshift for the three samples.
 All three flux ranges show a broader range in star-formation histories at lower redshifts, but the  $S_{\rm 870\mu m}$\,$=$\,0.5--1.5\,mJy galaxies at $z_{\rm med}$\,$\ls$\,2 have the largest intrinsic dispersion in star-formation histories, by $\sim$\,0.4\,dex, hinting at a wider variety of evolutionary paths.

\subsection{Galaxy structure and stability}

Figure~6 shows the available archival  JWST NIRCam imaging  of the faint $S_{\rm 870\mu m}$\,$=$\,0.5--1.5\,mJy galaxies, including the  faint ALMA-detected $K$-band selected sources from \citet{Tadaki20}.     The majority of the thumbnails use  the same combination of filters: F200W for the blue channel, F356W for green and F444W for red.   These are the same filters and scaling as used for the typically $S_{\rm 870\mu m}$--brighter  galaxies shown in  \citet{Gillman24} and so the morphologies of our faint sources can be easily compared to those shown in Figure~1 of \citet{Gillman24}.  The panels are labelled with the 870-$\mu$m flux densities of the sources and are ordered by decreasing  redshift, although we see no marked trend in morphology, colour or apparent size with redshift.

We   visually classified the galaxies using the same methodology and scheme as used by \citet{Gillman24}, classifying the sample into: major mergers (M), minor mergers or interactions (I), regular discs (D), or compact (C) sources.  For a source to be visually classified as a possible major merger relies on it being strongly disturbed, with tidal features, or a disturbed morphology with nearby companions that have similar brightnesses and colours, while potential minor mergers and interactions encompass less disturbed galaxies, with asymmetries or fainter nearby companions (see also \citealt{Gillman24}).  The relative fractions of each  class are:  major mergers, 17\,$\pm$\,7 per cent; minor mergers and interactions, 37\,$\pm$\,11 per cent; regular discs, 43\,$\pm$\,12 per cent; and compact, 3\,$\pm$\,3 per cent.    These ratios are statistically indistinguishable to those for the $S_{\rm 870\mu m}$\,$\geq$\,2\,mJy  sources in \citet{Gillman24}:  major mergers, 19\,$\pm$\,7 per cent;  minor mergers/interactions, 43\,$\pm$\,8 per cent; regular discs, 36\,$\pm$\,7 per cent; and compact, 2\,$\pm$\,2 per cent.    This indicates no strong variation in the external triggering mechanism of dusty star-forming galaxies with submillimetre flux density (see also \citealt{McKinney25,McKay25,Chan25}). Looking at the $z_{\rm med}$\,$=$\,1.5--2.5 and  $z_{\rm med}$\,$=$\,2.5--3.5 subsets we also see that the rates of major mergers, minor mergers and interactions and regular discs are consistent between the two redshift ranges.

One  visual distinction between the faint and bright submillimetre sources is the higher rate  of bright compact emission
components in the fainter population:  we identify such components in 54\,$\pm$\,9 per cent of the $S_{\rm 870\mu m}$\,$\geq$\,2\,mJy sources in \citet{Gillman24}, compared to 77\,$\pm$\,16 per cent in our faint sample -- although this difference is not statistically significant. To provide a more quantitative comparison we  used  S\'ersic fits to the F444W imaging of the faint submillimetre galaxies.  The imaging and methods are identical to those employed by \citet{Gillman24} in their study of a sample of typically-brighter submillimetre galaxies and we refer the reader to that work for more details of the fitting.
We find that the faint submillimetre sources have a  smaller $R_e$ in the F444W band and exhibit a higher S\'ersic $n$ (supporting the visual impression of more compact emission): $\langle R_{e}^{444}\rangle$\,$=$\,1.83\,$\pm$\,0.26\,kpc and $\langle n^{444}\rangle$\,$=$\,1.43\,$\pm$\,0.22
compared to $\langle R_{e}^{444}\rangle$\,$=$\,2.42\,$\pm$\,0.20\,kpc and $\langle n^{444}\rangle$\,$=$\,1.26\,$\pm$\,0.12
for the brighter sources,  but  the significance of these differences is only $\sim$\,2\,$\sigma$.   Within the faint sample, the lower-redshift subset ($z_{\rm med}$\,$=$\,1.5--2.5) show smaller $R_e$ and higher $n$ than the higher-redshift sources ($z_{\rm med}$\,$=$\,2.5--3.5): $\langle R_{e}^{444}\rangle$\,$=$\,1.82\,$\pm$\,0.27\,kpc and  $\langle n^{444}\rangle$\,$=$\,1.51\,$\pm$\,0.26, versus $\langle R_{e}^{444}\rangle$\,$=$\,3.24\,$\pm$\,0.88\,kpc and  $\langle n^{444}\rangle$\,$=$\,0.71\,$\pm$\,0.13, although yet again these differences are not statistically significant.

We can similarly compare the  dust continuum sizes of the  obscured activity within the populations using the effective radii, $R_e$, measured at 870-$\mu$m reported in \citet{Tadaki20} and \citet{Gullberg19}.  These sizes were derived from fitting S\'ersic $n$\,$=$\,1 models to the ALMA $uv$ visibility data.  For the faint $S_{\rm 870\mu m}$\,$=$\,0.5--1.5\,mJy galaxies with dust continuum sizes, all of which are at $z$\,$\leq$\,2.5,  we obtain a median size of $R_e^{\rm 870\mu m}$\,$=$\,2.1\,$\pm$\,0.4\,kpc (see also \citealt{Tadaki20}), which is consistent with their F444W sizes (they have $R_e^{\rm 870\mu m}/R_e^{444}$\,$=$\,1.1\,$\pm$\,0.3) and somewhat larger than the $R_e^{\rm 870\mu m}$\,$=$\,1.2\,$\pm$\,0.4\,kpc reported for the brighter $S_{\rm 870\mu m}$\,$\geq$\,2\,mJy sources from \citet{Gullberg19}  (which have $R_e^{\rm 870\mu m}/R_e^{444}$\,$=$\,0.4\,$\pm$\,0.1, \citealt{Gillman24}), although still statistically consistent within the errors. This offset is in line with the structural model suggested by \citet{Gullberg19} (and the dust-size behaviour reported by \citealt{Wang24}), whereby the dust continuum emission in brighter submillimetre sources is increasingly dominated by a bright, very compact component superimposed on fainter emission arising from dust in an extended component with a size similar to the stellar discs.

%
%
\begin{figure*}
    \centerline{         \psfig{file=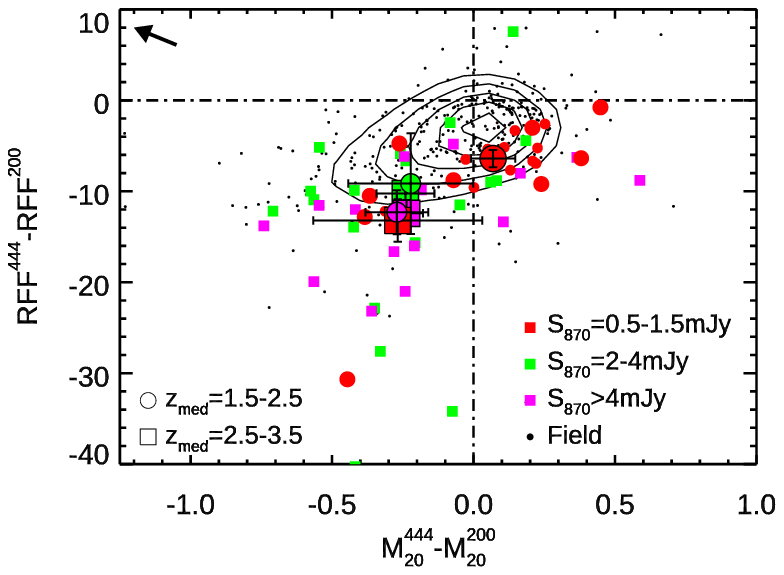, width=5in, angle=0} }
    \caption{\small
     The distribution of the difference in the  M$_{20}$ measurements between the F444W and F200W bands and the corresponding differences in RFF
     between F444W and F200W, for the $S_{\rm 870\mu m}$\,$=$\,0.5--1.5\,mJy, $S_{\rm 870\mu m}$\,$=$\,2--4\,mJy and $S_{\rm 870\mu m}$\,$\geq$\,4\,mJy samples at $z_{\rm med}$\,$=$\,1.5--2.5 and $z_{\rm med}$\,$=$\,2.5--3.5, following \citet{Gillman24}.  The faint sample combines  AS2UDSx,  AS2UDS and \citet{Tadaki20} sources, with those from \citet{Tadaki20} identified by small symbols.
  For comparison we show the distribution of massive star-forming field population at $z_{\rm med}$\,$=$\,1.5--2.5 from \citet{Gillman24} with a median $\langle$\,SFR\,$\rangle$ \,$\sim$\,25\,M$_\odot$\,yr$^{-1}$ and their number density indicated by contours.  
Larger symbols with error bars show the median values for each sample and their bootstrap errors.  These indicate that the faint sources at $z_{\rm med}$\,$\leq$\,2.5 have  morphological properties that more closely match the less-active, massive $z_{\rm med}$\,$=$\,1.5--2.5   field population, while the higher-redshift faint sources and all of the brighter submillimetre sources show much stronger variations in structure with wavelength.  The  arrow in the top-left indicates the expected change in $\Delta$\,RFF and $\Delta$\,M$_{20}$ measurements when using F356W and F200W, rather than F444W and F200W (roughly corresponding to the same restframe sampling) for the $z_{\rm med}$\,$=$\,1.5--2.5 samples.}
\end{figure*}

To identify more detailed structural differences in the various submillimetre galaxy samples we next turned to the RFF$^{444}$\,$-$\,RFF$^{200}$ versus M$_{20}^{444}$\,$-$\,M$_{20}^{200}$ plot from \citet{Gillman24}, their Figure~11d. This plot shows the difference in the residuals from a S\'ersic fit in the F444W and F200W bands against the difference in the M$_{20}$ values measured in F444W and F200W (see \citet{Gillman24} for a detailed description of the calculation of these quantities).   Values of RFF, the residual flux fraction, in a single band are zero for a galaxy that is  described by a perfectly smooth S\'ersic model, while more positive values indicate galaxies with increasingly large deviations from the best-fit smooth model.  A galaxy exhibiting significantly more structure in its F200W image compared to F444W will thus show more negative values of RFF$^{444}$\,$-$\,RFF$^{200}$.  M$_{20}$ measures the second moment of the brightest 20 per cent of pixels in the galaxy, normalised by the total moment for all pixels. Highly negative values of M$_{20}$ indicate clumpy light distribution, although the clumps are not necessarily at the centre of the galaxy \citep{Lotz04}.    Negative values of M$_{20}^{444}$\,$-$\,M$_{20}^{200}$ thus indicate that the light distribution in the galaxy appears clumpier in the F200W band than in F444W.

The  RFF$^{444}$\,$-$\,RFF$^{200}$ versus M$_{20}^{444}$\,$-$\,M$_{20}^{200}$  plot was the strongest diagnostic of the differences between typical submillimetre galaxies and a less-active, mass- and redshift-matched field sample in the analysis of \citet{Gillman24}.  They concluded that this was a result of strong structured dust obscuration within the submillimetre galaxies, compared to the less-active field population at similar redshifts and stellar masses.

Figure~7 shows the difference in the RFF and M20 measurements in the F444W and F200W bands  for the different submillimetre samples, including a comparison massive field sample at $z_{\rm med}$\,$=$\,1.5--2.5 taken from \citet{Gillman24}.    We see that the high-redshift faint and the intermediate and bright samples of submillimetre galaxies at all redshifts show offsets relative to both the field sample and the lower-redshift $S_{\rm 870\mu m}$\,$=$\,0.5--1.5\,mJy submillimetre galaxy sample (which lies close to the peak of the field population).  We find a  a 4.2-$\sigma$ offset between the the $S_{\rm 870\mu m}$\,$=$\,0.5--1.5\,mJy galaxies at $z=$\,1.5--2.5  and the equivalent flux-selected sample at $z_{\rm med}$\,$=$\,2.5--3.5  sources and 5.3-$\sigma$ between the $z_{\rm med}$\,$=$\,1.5--2.5 $S_{\rm 870\mu m}$\,$=$\,0.5--1.5\,mJy and the brighter samples at $z_{\rm med}$\,$=$\,1.5--3.5 (restricting these to  $z_{\rm med}$\,$=$\,1.5--2.5 gives a 3.5-$\sigma$ difference, primarily due to the larger uncertainties in the median values from the smaller samples).   We also indicate on Figure~7 the expected shift in the  $\Delta$\,RFF and $\Delta$\,M$_{20}$ measurements when these use F356W and F150W for the $z_{\rm med}$\,$=$\,1.5--2.5 samples (corresponding to similar restframe sampling to F444W and F200W for the sources at $z_{\rm med}$\,$=$\,2.5--3.5).  We conclude that any $K$-correction differences have a negligible impact on  the offsets between the  samples.

Figure~7 thus appears to support a wavelength-dependent morphological distinction between the $S_{\rm 870\mu m}$\,$=$\,0.5--1.5\,mJy galaxies at $z_{\rm med}$\,$=$\,1.5--2.5 and both the similarly faint sources at $z_{\rm med}$\,$=$\,2.5--3.5  and the brighter populations at similar and higher redshifts.   This distinction is in the sense that the faint, lower-redshift submillimetre sources more closely resemble the less-active field population: showing less structure and less variation in the light distributions in F200W (restframe $\sim$\,0.7\,$\mu$m) relative to F444W (restframe $\sim$\,1.5\,$\mu$m) compared to the $S_{\rm 870\mu m}$\,$\sim$\,1\,mJy galaxies at $z_{\rm med}$\,$\geq$\,2.5 and all of the brighter subsets.  Using high-resolution 870-$\mu$m ALMA observations, \citet{Gillman24} were able to directly link the F444W-bright, but red, pixels responsible for the offsets in the M$_{20}^{444}$\,$-$\,M$_{20}^{200}$ measurements with regions with high 870-$\mu$m surface brightness, indicating high dust column densities.  Our results suggest such clumpy, highly obscured regions (likely linked to the presence of compact, submillimetre-bright components) are present in
much of the submillimetre galaxy population, but may be absent in the  fainter, lower-redshift submillimetre sources (even though these systems may have slightly higher average $A_V$ than some of the other samples we have studied, see Table~2).

%
%
\begin{figure*}
    \centerline{         \psfig{file=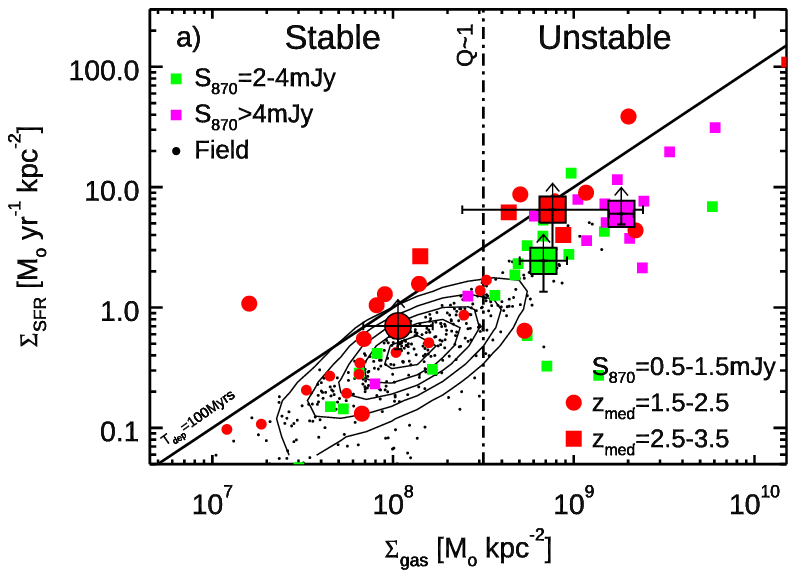, width=3.5in, angle=0}   \psfig{file=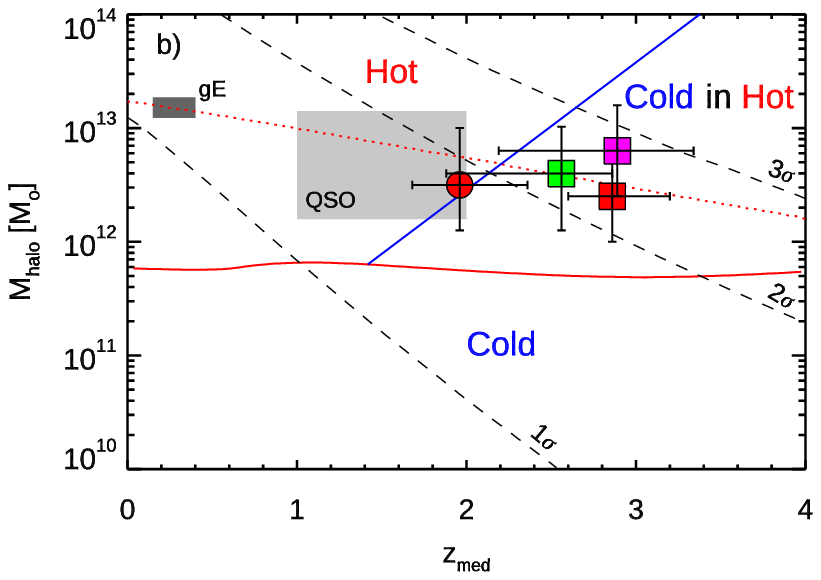, width=3.5in, angle=0} }
    \caption{\small      
   a) The same samples as Figure~7 plotted in gas and star-formation surface densities ($\Sigma_{\rm gas}$, $\Sigma_{\rm SFR}$) following \citet{Gillman24}. Splitting the faintest sample at $z_{\rm med}$\,$=$\,2.5, we find low-redshift faint sources have gas surface densities consistent with stable discs, similar to the less-active field population (contours). In contrast, high-redshift faint sources and all brighter submillimetre galaxies at $z_{\rm med}$\,$=$\,1.5--3.5 appear globally gravitationally unstable. A nominal Toomre $Q$\,$\sim$\,1 division  \citep{Gillman24} is shown, along with field galaxy distributions and a $\sim$\,100\,Myr depletion timescale track tracing the upper boundary on $\Sigma_{\rm SFR}$ at fixed $\Sigma_{\rm gas}$. The median $\Sigma_{\rm SFR}$ values are shown as lower limits, as the far-infrared sizes are typically smaller than F444W sizes, which are more comparable to those measured from low/mid-$J$ CO transitions (e.g., \citealt{Tadaki23,Rybak25}; Westoby et al., submitted).
   b) Estimated halo masses for flux-selected samples (Table~2) following \citet{Stach21}, compared to the \citet{Dekel06} model for gas accretion. Below the solid horizontal line, discs are fed by cold gas streams; above the solid diagonal line, these streams cannot penetrate hot haloes, halting the efficient  supply of cold gas. For the faint sample, halo masses at median redshifts ($z_{\rm med}$\,$=$\,1.5--2.5 and 2.5--3.5; error bars show 16--84$^{\rm th}$ ranges) suggest low-redshift faint galaxies lie in a regime with inefficient accretion of  cold gas, unlike higher-redshift and brighter systems. Halo masses are interpolated from \citet{Stach21} using median redshifts and baryonic masses. The dotted line shows median halo growth \citep{Fakhouri10}; shaded regions indicate halo masses for QSOs \citep{Eltvedt24} and massive ellipticals \citep{Sonnenfeld22}. Dashed lines show Press--Schechter collapse probabilities.
}
\end{figure*}

In Figure~8a  we plot the  samples from Figure~7 in terms of their gas and star-formation surface densities ($\Sigma_{\rm gas}$ and $\Sigma_{\rm SFR}$ respectively) following \citet{Gillman24}, using their integrated star-formation rates and dust masses (assuming half lies within the corresponding $R_e$) and our adopted $\delta_{\rm GDR}$ (see \S2.2).     We employ the restframe near-infrared sizes of the galaxies from JWST/NIRCam  to estimate the surface densities of the star formation and gas reservoirs in these galaxies.   Based on the comparisons of the 870-$\mu$m and F444W sizes of the submillimetre galaxies above (see also \citealt{Gillman24,Chen22, Hodge25}), we take  the F444W sizes as upper-limits to the restframe far-infrared sizes of the galaxies,  tracing the obscured majority of the star formation, giving  lower limits on $\Sigma_{\rm SFR}$.  We adopt the F444W size as a rough proxy for the low/mid-$J$ CO sizes of the galaxies to derive $\Sigma_{\rm gas}$  based on the results from \citet{Tadaki23}, \citet{Rybak25} and Westoby et al.\ (submitted).  The resulting median estimates of $\Sigma_{\rm gas}$ and $\Sigma_{\rm SFR}$ for the various flux-limited samples,   dividing the faintest samples at $z_{\rm med}$\,$=$\,2.5, are plotted in Figure~8a.

Figure~8a shows that the brighter submillimetre sources and also the higher-redshift   $S_{\rm 870\mu m}$\,$=$\,0.5--1.5\,mJy galaxies have comparable limits on the star-formation rate surface densities and also all exhibit relatively high  gas surface densities. These contrast with  the lower-redshift  $S_{\rm 870\mu m}$\,$=$\,0.5--1.5\,mJy galaxies and the bulk of the field population which have gas  surface densities that are $\sim$\,5--10\,$\times$ lower.  Following \citet{Gillman24} we illustrate the approximate location of the Toomre $Q$ parameter \citep{Toomre64}  $Q$\,$\sim$\,1 based on  the typical properties of the populations,  which indicates the rough boundary between gas discs that are stable to perturbations and those  that are unstable and so prone to fragmentation if  perturbed.   This boundary falls between the median gas surface density values of the two groups of submillimetre sources and suggests that the gas reservoirs in the lower redshift and fainter submillimetre galaxies are potentially stable (as are the bulk of the less-active field population at similar redshifts), whereas those with similar submillimetre flux densities at higher redshifts, $z_{\rm med}$\,$\geq$\,2.5, and the brighter submillimetre populations are all unstable systems.

\subsection{Halo masses and gas inflows}

To  understand the wider environment of the various submillimetre galaxy samples, and the potential impact of this on the supply of gas into the galaxies, we  turned to their halo masses.   As submillimetre galaxies have  significant stellar and  gas masses, we expect that the vast majority  are likely to be the central galaxies in their halos, although at the lowest redshifts, $z$\,$\sim$\,1.5, and lower masses it is possible that some  are satellites in larger halos \cite{Smail24}.

We attempted to gain  insight into the halo masses for the three flux-limited submillimetre samples from the clustering analysis of AS2UDS presented in \citet{Stach21} (see also \citealt{Hickox12,Chen16,Wilkinson17,Amvrosiadis19,Lim20}. That paper analysed the  AS2UDS  catalogue to derive halo masses for the  submillimetre galaxy samples in three redshift ranges $z_{\rm med}$\,$=$\,1.5--2.0, $z_{\rm med}$\,$=$\,2.0--2.5 and $z_{\rm med}$\,$=$\,2.5--3.0 (median redshifts of $\langle z_{\rm med}\rangle = 1.68\pm 0.02$, $\langle z_{\rm med}\rangle = 2.29\pm 0.03$ and $\langle z_{\rm med}\rangle = 2.69\pm 0.02$ respectively).   Using the  $S_{\rm 870\mu m}$--$z_{\rm med}$ trend in submillimetre galaxies we  very crudely estimated the likely halo masses for our three flux-selected samples by linearly interpolating the results from \citet{Stach21} to the median redshifts of our three samples (Table~2) and then scaling the halo masses using their  total baryonic masses (see Table~2) compared to those for the \citet{Stach21} samples:  $\langle M_{\rm bar}\rangle = (1.8\pm 0.2)\times 10^{11} $\,M$_\odot$, $\langle M_{\rm bar}\rangle = (2.0\pm 0.2)\times 10^{11} $\,M$_\odot$ and $\langle M_{\rm bar}\rangle= (1.7\pm 0.2)\times 10^{11} $\,M$_\odot$ for $z_{\rm med}$\,$=$\,1.5--2.0, $z_{\rm med}$\,$=$\,2.0--2.5 and $z_{\rm med}$\,$=$\,2.5--3.0 respectively, using our adopted dust-to-gas ratios.  This rescaling assumes a fixed baryon fraction within the range of these small corrections (typically $\sim$\,0.1--0.2\,dex) and we include a 0.3 dex error in quadrature with the halo mass errors to reflect the uncertainties of this procedure.   The  estimated halo masses are listed in Table~2, but we caution that given the rough calculation which these are based on they should be viewed as only indicative.

The estimated characteristic halo masses are shown in Figure~8b in comparison  to the schematic model for the thermal properties of accreting halos proposed by  \citet{Dekel06}. This model divides the halo mass--redshift plane into discrete regions based on the expected theoretical behaviour of cold gas accretion.   At the lowest halo masses, below $\sim$\,5\,$\times$\,10$^{11}$\,M$_\odot$ (galaxy baryonic masses of $\sim$\,5\,$\times$\,10$^{9}$\,M$_\odot$) the accretion of cold gas from the surrounding intergalactic medium onto galaxies, through streams and mergers with minor gas-rich satellites, continues at all redshifts down to the present day and so it can continue to efficiently fuel on-going star formation (this regime is labelled as ``Cold'' in Figure~8b).  In contrast, at higher  masses the infalling gas is expected to be shock heated to create a hot halo through which cold gas streams and gas-rich satellites can still penetrate until a critical redshift where this accretion can no longer directly fuel the central galaxy.  This transition is shown by the diagonal line, dividing the ``cold-in-hot'' regime (where the cold gas can reach the galaxy and replenish on-going star formation) and the ``hot'' regime where it cannot and so the gas reservoirs in the galaxies are expected to be increasingly depleted by  on-going star formation.

Our crude estimates of the halo masses for the various samples suggest that the bulk of the brighter submillimetre galaxies ($S_{\rm 870\mu m}$\,$\geq$\,2\,mJy) and the high-redshift end of the $S_{\rm 870\mu m}$\,$=$\,0.5--1.5\,mJy population likely fall in the regime where the massive submillimetre galaxies can continue to efficiently accrete cold gas from their surroundings and hence support their strong on-going star formation, focused on a highly-obscured kpc-scale central starburst visible in the submillimetre \citep{Gullberg19}.  In contrast, the $S_{\rm 870\mu m}$\,$=$\,0.5--1.5\,mJy sources at $z_{\rm med}$\,$\leq$\,2.5 fall on or below the threshold where efficient gas accretion from the wider environment is likely to be suppressed.   This hints at one possible explanation for the lower gas fractions and the resulting greater stability of the faint $S_{\rm 870\mu m}$\,$=$\,0.5--1.5\,mJy  sources at $z_{\rm med}$\,$\leq$\,2.5 -- that it is due to a shutdown of the flow of cold gas from the wider environment of these galaxies, which can then no longer support their high star-formation rates and feed a compact, obscured starburst.

\section{Discussion}

We begin by reiterating one critical point about the samples of ALMA-identified submillimetre galaxies that we have constructed in the UKIDSS UDS field.  This is that the number of faint sources, $S_{\rm 870\mu m}$\,$\ls$\,2\,mJy,  in the combined AS2UDSx+AS2UDS catalogue  is still severely under-represented compared to their true number in this region.  The combined catalogue includes roughly the expected number of $S_{\rm 870\mu m}$\,$\gs$\,2–-4\,mJy sources, but only $\sim$\,0.6 per cent of the $S_{\rm 870\mu m}$\,$\sim$\,1\,mJy sources lying within this area.  This imbalance needs to be borne in mind when discussing the trends in the population. 

The recovery of more faint submillimetre sources in the existing AS2UDS ALMA maps has increased the fraction of maps that host more than one submillimetre source.  We estimated that 36\,$\pm$\,2 per cent, or $\sim$\,280 of 791 sources, in the combined AS2UDSx+AS2UDS catalogue lie in a field containing another submillimetre galaxy,  However, we have estimated from the blank field counts that roughly one in four maps are expected to contain a  $S_{\rm 870\mu m}$\,$\gs$\,1\,mJy source by random chance, amounting to $\sim$\,180  sources.   Removing this contribution to yield a ``field--subtracted'' sample suggests an excess of around $\sim$\,100 of the remaining $\sim$\,610 sources are likely to be present due to clustering of faint submillimetre sources around  brighter  sources.   Thus  clustering is responsible for 36\,$\pm$\,4 per cent of the multiples, similar to the $\gs$\,30 per cent estimated by \citet{Stach18}, with the remaining two thirds are due to unrelated line-of-sight projections.    This dilution of the clustered examples of multiple submillimetre sources is the reason that our estimated surface density of $S_{\rm 870\mu m}$\,$\gs$\,1\,mJy sources is consistent with that derived from blank field measurements.

Considering trends in the sample with  870-$\mu$m flux density, we found that the variation of median redshift with 870-$\mu$m flux density had the form: $S_{\rm 870\mu m} = (0.40\pm 0.12) \times  (1+z_{\rm med})^{1.7\pm 0.2}$.   This behaviour can be understood from the relation of  $S_{\rm 870\mu m}$  with far-infrared luminosity and dust mass from Ikarashi et al.\ (in prep.): 
$S_{\rm 870\mu m} \propto M_{\rm d}^{0.68} \times L_{\rm IR}^{0.26}$.   The redshift evolution of $M_{\rm d}$ for the AS2UDS sample from \citet{Dudzeviciute20}, is $M_{\rm d} = (1.9\pm 0.3 \times  10^8) \times (1+z)^{0.96\pm 0.13}$, while the expected evolution of  $L_{\rm IR}$ is $\sim (1+z)^\alpha$ with $\alpha\sim 4$ once selection effects are considered \citep[e.g.,][]{Dudzeviciute21}.  These predict that the evolution in   $S_{\rm 870\mu m}$ will scale as $(1+z)^{\gamma_{L_{\rm IR}}+\gamma_{M_{\rm d}}}$ with $\gamma_{M_{\rm d}} \sim 0.68 \times 0.96 \sim 0.7$ and $\gamma_{L_{\rm IR}} \sim 0.26\times4 \sim 1.0$, matching the evolution of $\sim (1+z)^{1.7}$, as seen.   Thus, while 870-$\mu$m flux density predominantly tracks the dust mass of galaxies, the redshift evolution of $S_{\rm 870\mu m} $ more reflects   the strong evolution in characteristic far-infrared luminosity (and by extension star-formation efficiency) within the population.

More generally, we found that the faintest submillimetre galaxies, with $S_{\rm 870\mu m}$\,$\sim$\,1\,mJy, have physical  properties that broadly follow the trends seen for  brighter submillimetre sources (Figure~4).   The faint sources lie at  lower redshifts than the brighter population  and have slightly lower stellar masses, with lower dust-to-stellar ratios and inferred gas depletion timescales.   These characteristics of the faint population suggested they are more ``evolved'' on average than the typically higher-redshift, brighter sources, having less gas available for future star formation.   This difference became more marked when the samples were divided in redshift, with the $S_{\rm 870\mu m}$\,$\sim$\,1\,mJy sources at  $z_{\rm med}$\,$\ls$\,2.5 exhibiting increasingly distinct properties from either the higher-redshift sources at these flux densities, or the brighter populations,   with  lower gas fractions, shorter gas consumption timescales and potentially more diverse star-formation histories (Figure~5).

To investigate the origin of these differences we studied  the  structural properties  of those faint submillimetre sources that have archival high-resolution near-infrared imaging from  JWST NIRCam  (Figure~6).  We contrasted the properties of these faint sources with those derived from an identical analysis by \citet{Gillman24} of typical brighter submillimetre galaxies, many of them from the AS2UDS catalogue. A visual classification of the morphologies of the faint sources suggested that 17\,$\pm$\,7 per cent are potential major mergers, while 37\,$\pm$\,11 per cent could be minor mergers or interactions, with the bulk of the remainder appearing to be regular discs.      These proportions are statistically indistinguishable to those for the $S_{\rm 870\mu m}$\,$\geq$\,2\,mJy  sources in \citet{Gillman24}, showing that there is no strong variation in the external triggering mechanism of dusty star-forming galaxies with submillimetre flux density (e.g., \citealt{McKinney25,McKay25,Chan25}).

A quantitative analysis of the  near-infrared structures of the galaxies provided hints of differences between the samples, with the fainter sources having higher S\'ersic $n^{444}$ and smaller effective radii in the F444W band, $R_e^{444}$, than the brighter submillimetre sources,  although they  exhibited larger dust continuum sizes at 870\,$\mu$m: $R_e^{\rm 870\mu m}$\,$=$\,2.1\,$\pm$\,0.4\,kpc (as noted by \citealt{Tadaki20}) compared to   $R_e^{\rm 870\mu m}$\,$=$\,1.2\,$\pm$\,0.4\,kpc, and corresponding higher ratios of dust to F444W sizes: $R_e^{\rm 870\mu m}/R_e^{444}$\,$=$\,1.1\,$\pm$\,0.3 versus $R_e^{\rm 870\mu m}/R_e^{444}$\,$=$\,0.4\,$\pm$\,0.1. Moreover,  within the faint sample, the low redshift subset had the higher S\'ersic $n^{444}$ and smaller $R_e^{444}$, although none of these differences were statistically significant.

However, one strong morphological distinction was found between the $S_{\rm 870\mu m}$\,$\sim$\,1\,mJy galaxies at $z_{\rm med}$\,$\ls$\,2.5 and both the similarly faint sources at $z_{\rm med}$\,$\gs$\,2.5  and the brighter populations at comparable and higher redshifts.  This was when comparing quantitative morphological measures that are sensitive to  structured dust obscuration:  the difference between  RFF and M$_{20}$ values measured in the F200W and F444W passbands (Figure~7).      \citet{Gillman24} had shown that these parameters  distinguish typical bright submillimetre galaxies from less active, but similarly massive star-forming galaxies at the same redshifts.    Our analysis showed that this distinction from the field population applied not only to the brighter submillimetre populations,  those with $S_{\rm 870\mu m}$\,$\geq$\,2\,mJy, but also those  $S_{\rm 870\mu m}$\,$\sim$\,1\,mJy sources at $z_{\rm med}$\,$\gs$\,2.5.

In contrast, the $S_{\rm 870\mu m}$\,$\sim$\,1\,mJy sources at $z_{\rm med}$\,$\ls$\,2.5 had much weaker signatures of structured dust obscuration and were much more similar to the bulk of the less-active, massive field population.     This behaviour, along with size trends in the stellar and dust emission, are consistent with the structural model from \citet{Gullberg19} where by the dust continuum emission from these massive galaxies comprise a faint extended component associated with star-formation activity in the  stellar disc, with a characteristic flux density of $S_{\rm 870\mu m}$\,$\sim$\,0.5\,mJy, superimposed upon which is a more compact component, $\sim$\,1\,kpc, that dominates in the brighter submillimetre sources.   \citet{Gullberg19} proposed that this brighter, compact emission arises from a central dust-obscured starburst linked to the formation and growth of a spheroid or bulge (it is also related to the so-called ``starbursts-within-the-main-sequence'' discussed by \citealt{Elbaz18,Puglisi21} and \citealt{GomezGuijarro22b}).  The absence of this compact dust component in the  $S_{\rm 870\mu m}$\,$\sim$\,1\,mJy population at $z$\,$\ls$\,2.5 may be because they have already grown a more significant stellar bulge component, as hinted at by the morphological analysis in \S3.3 and discussed further below, which stabilises their gas discs and suppresses gas flows into the galaxy centres.

To understand the relationship between this model and the  absence of structured dust obscuration in the fainter and lower redshift submillimetre galaxies, we followed \citet{Gillman24} and  compared the inferred gas and star-formation rate densities in the different samples in Figure~8. This showed that the brighter submillimetre sources and also the higher-redshift faint galaxies with $S_{\rm 870\mu m}$\,$\sim$\,1\,mJy  all exhibited  high  gas surface densities, in contrast  the lower-redshift  $S_{\rm 870\mu m}$\,$\sim$\,1\,mJy  systems had gas  surface densities that are roughly an order of magnitude lower and consistent the bulk of the field population.  The two sets of samples straddle the expected critical gas density corresponding to a Toomre parameter of $Q$\,$\sim$\,1 \citep{Toomre64}, which is the gas density threshold above which gas discs become  unstable and  prone to fragmentation if  perturbed.   This implies that the gas reservoirs in the lower redshift and fainter submillimetre galaxies are potentially stable (similar to the less-active field population), whereas those within $S_{\rm 870\mu m}$\,$\sim$\,1\,mJy sources at higher redshifts, $z_{\rm med}$\,$\gs$\,2.5, and the brighter submillimetre populations are all unstable.    This  suggests a fundamental difference between the  galaxies with $S_{\rm 870\mu m}$\,$\sim$\,1\,mJy at  $z_{\rm med}$\,$\ls$\,2.5 and the other submillimetre populations related to their gas content.

Finally, we investigated the potential differences in the environments of the submillimetre galaxy samples and the influence that this may have on their gas accretion.   This analysis exploited the trend between submillimetre flux density and redshift to attempt to crudely estimate the expected halo masses of the different samples by a rough interpolation of the results from the clustering analysis of the AS2UDS sample in \citet{Stach21}.    When compared to the schematic model for accreting halos proposed by  \citet{Dekel06} the estimated halo masses of the $S_{\rm 870\mu m}$\,$\sim$\,1\,mJy galaxies at  $z_{\rm med}$\,$\ls$\,2.5 fall on or below the threshold in  the model where efficient cold gas accretion from the wider environment is likely to be suppressed due to the formation of a halo of hot gas around the galaxy,   in this situation the existing gas reservoirs in the galaxies can be depleted by rapid on-going star formation.  In contrast, both the  higher-redshift $S_{\rm 870\mu m}$\,$\sim$\,1\,mJy galaxies and the brighter submillimetre sources had estimated halo masses and median redshifts that placed them in a regime where the model indicated that the efficient accretion of gas from cold streams and gas-rich satellites can still  support their on-going star formation.   This distinction suggests a possible cause for the lower gas fractions,  and the resulting greater stability and hence absence of central obscured, starbursts, in the faint $S_{\rm 870\mu m}$\,$\sim$\,1\,mJy  sources at $z_{\rm med}$\,$\ls$\,2.5 that is driven by the interruption of the flow of cold gas from their wider environments, so that they can then no longer support their high star-formation rates.

Taken together the structural and environmental analyses suggested that it is not the dynamical perturbations due to mergers which trigger submillimetre activity in these galaxies \citep[see][]{McAlpine19}, but rather that the intense obscured star formation activity requires the  availability of a  massive  and unstable gas reservoir within the galaxy, which can be triggered to fragment by  interactions or through secular processes in  undisturbed discs.   For the fainter and lower redshift submillimetre sources the restriction of their accretion of cold gas driven by their surrounding hot halos limits their gas fractions and  results in gas discs that are more stable,   and potentially further stabilised by  more significant bulge components, which in turn  appears to be responsible for the reduced signatures of structured dust obscuration across the sources and the lack of compact central obscured starbursts.

\subsection{Evolutionary links}

Beyond investigating the internal and external processes that are responsible for the dust-obscured activity in  the submillimetre population, we are also interested in the evolutionary history of these galaxies and their connection to  populations at lower redshifts.

Our analysis suggests that  faint submillimetre galaxies ($S_{\rm 870\mu m}$\,$\sim$\,1\,mJy) at $z$\,$\gs$\,2.5 share many characteristics with the better-studied  bright submillimetre population.  However, as we have noted these sources are under-represented in our catalogue compared to their true surface density by around a factor of $\sim$\,150.    As a result these systems are expected to be the dominant precursors of any  descendant population that passed through a submillimetre-bright phase.  This in-balance in the relative sampling of the different 870-$\mu$m brightness sources needs to be accounted for when assessing the characteristics of their likely descendants.   Hence,  we estimate that the current median stellar masses of the full submillimetre population with $S_{\rm 870\mu m}$\,$\gs$\,1\,mJy is $\langle M_\ast \rangle$\,$=$\,(5.8\,$\pm$\,1.1$^{+5.9}_{-4.0}$)\,$\times$\,10$^{10}$\,M$_\odot$, with a gas mass of $\langle M_{\rm gas}\rangle$\,$=$\,(2.4\,$\pm$\,0.2$^{+1.6}_{-0.9}$)\,$\times$\,10$^{10}$\,M$_\odot$, meaning that their present-day descendants would have a modal stellar mass of $\langle M_\ast^{z=0} \rangle$\,$=$\,(8.6\,$\pm$\,0.7$^{+6.5}_{-3.0}$)\,$\times$\,10$^{10}$\,M$_\odot$ if their gas reservoirs were completely turned into stars and there was no subsequent accretion.   This compares to the modal stellar mass of $\sim$\,4$^{+2}_{-3}$\,$\times$\,10$^{10}$\,M$_\odot$ for field ellipticals and S0/Sa's  \citep{Kelvin14} and the quiescent cluster population at $z$\,$\sim$\,0 \citep{Park26}.

To investigate the broader evolution of these submillimetre samples,  Figure~8b shows the predicted  mass evolution of halos as a function of      redshift from the theoretical study of \citet{Fakhouri10}.  This suggests a  link between the estimated halo masses of the  submillimetre samples with that derived for   massive elliptical galaxies with $M_\ast$\,$\sim$\,4\,$\times$\,10$^{11}$\,M$_\odot$     from  \citet{Sonnenfeld22}. The crude estimates of the halo masses of the various submillimetre galaxy samples are broadly consistent with these galaxies  evolving into massive elliptical galaxies by $z$\,$\sim$\,0, as well as being potential progenitors of optically-bright QSOs \citep{Eltvedt24}, at $z$\,$\sim$\,1--2 (as previously proposed by \citealt{Hickox12}, amongst others).

Our results imply a distinction between the characteristics of the bright and faint submillimetre galaxies as a function of redshift, that suggest different evolution -- but  are these populations linked?  Comparing  the properties of the faint ($S_{\rm 870\mu m}$\,$\sim$\,1\,mJy) and brighter populations,  we found that the  two populations have similar stellar masses, similar star-formation rates, but with the fainter sources having lower inferred gas masses and typically  lower redshifts.    The $S_{\rm 870\mu m}$\,$\sim$\,1\,mJy galaxies are thus  not merely scaled-down versions of the brighter sources.   But could the faint submillimetre sources be linked through an evolutionary cycle to the brighter sources, could they be descendants of bright submillimetre sources caught as they are being quenched?

In general, No.  Most of the  faint sources are  not descendants of the bright sources:  the $\sim$\,150\,$\times$  difference in space densities precludes this as the visibility lifetimes from their relative  gas depletion timescales accounts for just $\sim$\,3\,$\times$ difference in space densities.   Hence, even at $z$\,$\gs$\,3 there are still too many $S_{\rm 870\mu m}$\,$\sim$\,1\,mJy sources relative to the bright $S_{\rm 870\mu m}$\,$\gs$\,2\,mJy population by a factor of $\sim$\,20 even after accounting for their relative lifetimes.   Thus while bright submillimetre sources may evolve through a faint  phase, most faint submillimetre galaxies aren't formed this way, even at high redshifts  where bright submillimetre galaxies are most prevalent.    

However, at the highest baryonic masses and highest redshifts, it may be the there is an evolutionary link between the faint and bright submillimetre galaxies.   Focusing just on those galaxies with baryonic masses of $M_{\rm bar}$\,$\geq$\,1\,$\times$\,10$^{11}$\,M$_\odot$ at $z$\,$\geq$\,2.5, this selection contains two thirds of the $S_{\rm 870\mu m}$\,$\geq$\,4\,mJy galaxies and half of the $S_{\rm 870\mu m}$\,$=$\,2--4\,mJy sample, compared to $\sim$\,10 per cent of all the $S_{\rm 870\mu m}$\,$\sim$\,1\,mJy sources (but $\sim$\,40 per cent of those at $z$\,$\geq$\,2.5).    Correcting for the sampling of the different populations in our sample, we estimate that $S_{\rm 870\mu m}$\,$\sim$\,1\,mJy sources at $z$\,$\geq$\,2.5 are only $\sim$\,5\,$\times$ more numerous than $S_{\rm 870\mu m}$\,$\gs$\,2\,mJy galaxies  at these redshifts.  The faint sources have similar stellar masses to the brighter galaxies, but lower gas masses ($\sim$\,4\,$\times$) and slightly lower star-formation rate ($\sim$\,2\,$\times$),  they lie at slightly lower redshifts ($z$\,$\sim$\,3.0 versus $z$\,$\sim$\,3.2, or $\sim$\,200\,Myrs), and have older inferred ages (by $\sim$\,300\,Myrs).  Given a dust removal timescale of $\sim$\,400\,Myrs at $z$\,$\sim$\,3 \citep{Lesniewska25}, all  of these characteristics are consistent with a model where bright high-redshift submillimetre sources evolve through a fainter submillimetre phase as their activity quenches, potentially becoming $z$\,$\sim$\,3 post-starburst or quiescent galaxies.  However,   this evolutionary path only contributes significantly to  the $S_{\rm 870\mu m}$\,$\sim$\,1\,mJy population for the most massive galaxies at higher redshifts.    As a result, the bulk of the population of faint submillimetre galaxies evolve independently of the brighter submillimetre sources, with those at $z$\,$\ls$\,2 being members of the dominant population of LIRGs,  with properties more similar to the typical less-active, but massive star-forming field population, including secularly-driven star formation in generally stable gas reservoirs.

We conclude that bright submillimetre galaxies are intrinsically rare events that require especially large masses of gas in the most massive halos:  they are rare in part perhaps because of the need for a deficit in previous activity, either star formation or AGN-driven -- to allow a massive reservoir of gas to build up \citep{McAlpine19}, or because an event such as a major merger brings the gas together for the first time.   In contrast, faint submillimetre galaxies at lower redshifts can form from somewhat less massive galaxies ($\sim$\,0.3 dex, which given the steep  mass function means  $\sim$\,3--5\,$\times$ more abundant), and somewhat less gas-rich systems.  These become more frequent at lower redshifts where the numbers of suitably massive galaxies (with more extended formation histories) is growing rapidly, although the suppression of gas cooling onto these central galaxies acts to  reduce their numbers below $z$\,$\sim$\,1--2.

\section{Conclusions}

The goal of this work was to increase the sample of faint submillimetre galaxies, with $S_{\rm 870\mu m}$\,$\sim$\,1\,mJy, in the well-studied UKIDSS UDS field and to then exploit these to investigate the variation in their properties as a function of submillimetre flux density to identify if there is a  duality in the star-formation processes seen in these highly dust obscured and strongly star-forming systems and that in  the less active field population, or whether they simply represent a continuum.  The main conclusions of this work are:

We revisited the cataloguing of submillimetre galaxies in the ALMA 870-$\mu$m maps in the UKIDSS UDS field from the AS2UDS survey \citep{Stach19}, constructing a catalogue of lower significance sources, down to 3.1\,$\sigma$, compared to the 4.3-$\sigma$ limit in the previous work.  To reduce the  contamination from false positive sources in the sample we matched the initial sample to galaxies uncovered in the very deep $K$-band imaging of this region ($K$\,$=$\,25.7).  However, we also needed to apply  colour and redshift cuts, $(H-K)$\,$\geq$\,0.4 and $z_{\rm med}$\,$\geq$\,1, to the $K$-band counterparts to reduce the false-positive rate in the final sample  to an acceptable level, $\sim$\,7 per cent.  We provide the AS2UDSx catalogue of 84 new sources in Table~1, where we expect six of these to be false-positives.  These galaxies have submillimetre flux densities of $S_{\rm 870\mu m}$\,$\sim$\,0.27--2.2\,mJy and they double the number of sources with $S_{\rm 870\mu m}$\,$\sim$\,1\,mJy compared to the original AS2UDS sample of \citet{Stach19}.  The combined faint AS2UDSx+AS2UDS sample is also roughly twice the size of the    available literature sample at these flux densities available from a heterogeneous mix of  archival compilations, lensing and blank field surveys at 870--1200\,$\mu$m.

We found that around half of the new faint submillimetre sources lie in AS2UDS maps that previous lacked a detected source in  the catalogue from \citet{Stach19}.   This provides circumstantial support that the many of the new sources are real submillimetre galaxies.   These new sources also increase the multiplicity in the combined AS2UDSx+AS2UDS sample, with 36\,$\pm$\,2 per cent of the submillimetre sources now lying in an ALMA map which contains another submillimetre galaxy.    

After correcting for the contamination and incompleteness in our sample from the $K$-band, colour and redshift selections, we  estimated the surface density of   $S_{\rm 870\mu m}$\,$\sim$\,1\,mJy sources in the combined AS2UDSx+AS2UDS sample to be  (2.2\,$\pm$\,0.3)\,$\times 10^4$ degrees$^{-2}$.  This is consistent with published sub/millimetre number counts of  (2\,$\pm$\,1)\,$\times$\,10$^4$ degrees$^{-2}$ derived from narrow blank-field or cluster lensing surveys with ALMA.   The absence of a strong excess in the  counts due to clustering of these faint submillimetre galaxies around the typically brighter submillimetre sources targeted by AS2UDS suggests that flux boosting as a result of source blending in low-resolution SCUBA-2 surveys is not a strong bias at flux densities of $S_{\rm 870\mu m}$\,$\sim$\,1\,mJy, likely because the source densities are  approaching the equivalent of one source per 14.5$''$ FWHM SCUBA-2 beam or $\sim$\,8\,$\times 10^4$ degrees$^{-2}$.

The faint sources in our new sample follow the trends with submillimetre flux density seen in brighter samples, with the $S_{\rm 870\mu m}$\,$\sim$\,1\,mJy galaxies lying at lower redshifts, having lower gas fractions and shorter gas depletion timescales, but similar specific star-formation rates and broadly similar stellar masses to the brighter submillimetre population.  However, when divided in redshift, the lower-redshift ($z_{\rm med}$\,$\ls$\,2) examples have more diverse properties that deviate more from the brighter populations.

We suggest that while the 870-$\mu$m brightness of galaxies predominantly reflects their dust mass, the reported trend of increasing median source redshift with 870-$\mu$m flux density  (e.g., \citealt{Stach19}) is driven by the strong evolution in characteristic far-infrared luminosity (and by extension star-formation efficiency) in the population at higher redshifts.

We used archival JWST NIRCam imaging of our expanded sample of faint submillimetre galaxies to search for morphological evidence of differences in the galaxy populations as a function of flux density.   We found comparable rates of visually identified mergers or interactions in the $S_{\rm 870\mu m}$\,$\sim$\,1\,mJy systems as derived for brighter samples by earlier studies using identical imaging, indicating no strong variation in the rate of potential external triggering of their activity with submillimetre flux density across $S_{\rm 870\mu m}$\,$\sim$\,1--10\,mJy.  Using quantitative morphological measure we found that the faintest and lower redshift ($z$\,$\ls$\,2) subset appear to be morphologically similar to the similarly massive, but less active star-forming field population at similar redshifts, and have morphological signatures of structured dust obscuration that are distinct from either the  $S_{\rm 870\mu m}$\,$\sim$\,1\,mJy galaxies at $z$\,$\gs$\,3 or the brighter and typically higher-redshift submillimetre galaxies.  

We investigated possible causes for this morphological difference and concluded that the $S_{\rm 870\mu m}$\,$\sim$\,1\,mJy population at $z_{\rm med}$\,$=$\,1.5--2.5 has inferred gas surface densities that are roughly an order of magnitude lower than those estimated for the higher redshift or brighter samples.   This  analysis suggested that the gas reservoirs in the higher redshift and brighter sources are likely to be globally unstable and so prone to bursts of intense star formation if perturbed through major or minor mergers, the accretion of gas or secular processes.  This activity would support the presence of highly dust-obscured kpc-scale activity in the central regions of these systems \citep{Gullberg19}, which have been linked to the formation of bulges and spheroids \citep{Hodge16,Hodge19,Hodge25}.  In contrast the $S_{\rm 870\mu m}$\,$\sim$\,1\,mJy sources at $z_{\rm med}$\,$\ls$\,2 appear more stable due to their lower inferred gas surface densities (and potentially higher existing bulge-to-disc ratios, that may further stabilise their gas reservoirs).  This may explain why the appear less likely to contain highly structured obscuring dust and bear a closer resemblance to less active, massive field galaxies at similar redshifts.     

We also employed published estimates of the halo masses of submillimetre galaxies from \cite{Stach21} to crudely estimate the halo masses for the various submillimetre samples.  Compared to the theoretical framework for gas accretion onto halos from \citet{Dekel06}, we   found that the median redshift and inferred  halo masses for the  $S_{\rm 870\mu m}$\,$\sim$\,1\,mJy population at $z$\,$\ls$\,2 place these galaxies in a regime where there is expected to be suppression of the  accretion of  cold gas from their surroundings, which would limit their star formation activity.  In contrast the $S_{\rm 870\mu m}$\,$\sim$\,1\,mJy sources at $z$\,$\gs$\,3 and the brighter submillimetre samples lie in a regime where the galaxies are expected to continue to efficiently accrete  gas through cold flows or minor mergers, enabling these systems to continue to replenish their gas reservoirs and so continue to support the vigorous star formation  seen in the obscured centres.

We conclude that the  brighter and higher redshift submillimetre galaxies are massive, strongly star-forming galaxies that are gas rich and  unstable systems, hosting obscured kpc-scale central starbursts fueled by mergers and  gas inflows from their wider environment.  However, at fainter submillimetre flux densities around $S_{\rm 870\mu m}$\,$\sim$\,1\,mJy, the much more numerous population exhibits a dichotomy in its properties.   The fainter submillimetre galaxies at $z$\,$\ls$\,2 have characteristics that are driven by declining gas accretion from their wider environment,  more diverse previous evolutionary pathways (including potentially more significant prior star formation and AGN activity which reduces the available gas reservoirs, as well as possibly contributing to higher bulge growth that stabilises their gas discs), more stable gas reservoirs and  shorter durations for the current levels of star formation due to the limited size of those reservoirs.  In essence these galaxies represent the most active members of the field population hosting secularly-driven star formation across their discs, and lack intense central starbursts.    In contrast the less numerous $S_{\rm 870\mu m}$\,$\sim$\,1\,mJy sources at $z$\,$\sim$\,3 are more similar to the brighter submillimetre population, with the potential for continued replenishing of their cold gas reservoirs via external accretion,  gas discs that are unstable (Toomre $Q$\,$\ls$\,1) and where the current star-formation event plays a more significant role in the  evolution of the galaxies.  A fraction of the latter sources may comprise the descendants of brighter submillimetre galaxies undergoing quenching, but we estimate that these contribute only a minority of the faint submillimetre galaxies at $z$\,$\sim$\,3.

In summary, we suggest that  submillimetre-selected galaxies comprise a mix of two distinct populations with a transition around $S_{\rm 870\mu m}$\,$\sim$\,1\,mJy and $z$\,$\sim$\,2.   At the bright end, $S_{\rm 870\mu m}$\,$\sim$\,2--20\,mJy, there is a population of high-redshift ($z$\,$\gs$\,2), unstable, gas-rich and  hence strongly star-forming, galaxies with a unique characteristic: most of their star-formation activity is occurring in a compact dust-obscured core, likely linked to  spheroid formation.  In contrast, the much more numerous sources at the faint end, $S_{\rm 870\mu m}$\,$\ls$\,1\,mJy, are less gas-rich, more evolved and less active  galaxies at $z$\,$\ls$\,2, where star formation is occurring across their extended discs.  The transition between these two populations is likely to be linked to the reported evolution in the flattening of the high-mass ``main sequence'' at $z$\,$\gs$1.5 \citep{Popesso23}.

\section*{Acknowledgements}

We thank Stuart Stach, James Simpson, Omar Almaini, Will Hartley, David Maltby, Chris Simpson, Vinod Arumugam and Rob Ivison for their work on the various surveys used in this study.   We also thank Soh Ikarashi and Jackie Hodge for discussions about topics related to this work.

I.R.S.\ and A.M.S.\ acknowledge support from STFC (ST/X001075/1).
S.R.G.\ and U.D.\ acknowledge funding from the Cosmic Dawn Center (DAWN) funded by the Danish National Research Foundation under grant DNRF140.

This paper made use of observations from  the following ALMA projects: ADS/JAO.ALMA\#2012.1.00090.S,
\#2015.1.01528.S, \#2016.1.00434.S and \#2017.1.01492.S.     ALMA is a partnership of ESO (representing its member states), NSF (USA), and NINS (Japan), together with NRC (Canada), NSC and ASIAA (Taiwan), and KASI (Republic of Korea), in cooperation with the Republic of Chile. The Joint ALMA Observatory is operated by ESO, AUI/NRAO, and NAOJ.  UKIDSS-DR11 photometry made use of UKIRT.  When the data reported here were acquired, UKIRT was operated by the Joint Astronomy Centre on behalf of the Science and Technology Facilities Council of the U.K.  This work is based in part on archival data obtained with the Spitzer Space Telescope and the NASA/IPAC Extragalactic Database (NED), which are operated by the Jet Propulsion Laboratory, California Institute of Technology under a contract with the National Aeronautics and Space Administration. This study has made use of NASA’s Astrophysics Data System Bibliographic Services.
This research has made use of the NASA/IPAC Infrared Science Archive, which is funded by the National Aeronautics and Space Administration and operated by the California Institute of Technology.  Extensive use was made of {\sc topcat} \citep{Taylor05} in this study. 

Facilities: ALMA, UKIRT, Subaru, CFHT, VISTA, Spitzer, Herschel, VLA, JWST.

\section*{Data availability}

All of the data used in this paper can be obtained from the ALMA, UKIRT, Subaru, CFHT, VISTA, Spitzer, Herschel, VLA and JWST data archives.

\bibliography{as2udsfaint}{}

\newpage

%
%

\setcounter{table}{0}

\begin{table*}
\caption{AS2UDSx 870-$\mu$m source catalogue.  Sources marked with $^\ast$ are the closest matches to the six negative sources that were selected using the same criteria applied to the positive sample presented here.  $^\dagger$ are potential AGN identified following \citet{Donley12}. }
\begin{tabular}{lcccp{0.3cm}ccccccc}
\hline \noalign {\smallskip}
ID & R.A.\ & Dec.\ & $S_{\rm 870\mu m}$ & $\!\!\!$SNR$_{0.5}$ &   R.A.$^{\! \! K}$ & Dec.$^{\!\! K}$ & $z_{\rm med}$ &  $M_{\ast}$ & SFR &  $M_{\rm d}$  & $A_V$ \\
  &  \multicolumn{2}{c}{(J2000)} & (mJy) & & \multicolumn{2}{c}{(J2000)} & & (10$^{11}$\,M$_\odot$) & (M$_\odot$\,yr$^{-1}$) &  (10$^{8}$\,M$_\odot$) &  \\
\hline \noalign {\smallskip}
AS2UDSx0030.14 & 34.16857 & $-$5.2273 & 0.27$\pm$0.12 & 3.2 & 34.16859 & $-$5.2273 & 3.50$\pm$0.01 &  1.10$\pm$0.01 &    159$\pm$1 &  0.33$\pm$0.14 & 0.69$\pm$0.01 \\ 
AS2UDSx0049.12 & 34.59358 & $-$5.4969 & 0.96$\pm$0.18 & 3.4 & 34.59347 & $-$5.4969 & 2.56$\pm$0.06 &  0.71$\pm$0.09 &    220$\pm$40 &  1.14$\pm$0.27 & 2.41$\pm$0.10 \\ 
AS2UDSx0054.23 & 34.88674 & $-$5.0774 & 0.89$\pm$0.18 & 3.2 & 34.88676 & $-$5.0774 & 3.17$\pm$0.13 &  0.51$\pm$0.08 &     40$\pm$20 &  2.15$\pm$0.97 & 0.99$\pm$0.25 \\ 
AS2UDSx0055.19 & 34.37463 & $-$5.0562 & 1.36$\pm$0.22 & 4.0 & 34.37476 & $-$5.0562 & 1.95$\pm$0.53 &  0.32$\pm$0.15 &    700$\pm$600 &  2.36$\pm$0.64 & 2.71$\pm$0.51 \\ 
AS2UDSx0057.32 & 34.81688 & $-$5.3017 & 1.51$\pm$0.27 & 3.1 & 34.81712 & $-$5.3016 & 1.88$\pm$0.17 &  1.32$\pm$0.15 &     15$\pm$7 &  4.14$\pm$1.25 & 1.04$\pm$0.46 \\ 
AS2UDSx0061.20$^\ast$ & 34.21198 & $-$5.2498 & 1.45$\pm$0.24 & 3.2 & 34.21185 & $-$5.2499 & 3.81$\pm$0.16 &  0.11$\pm$0.03 &     90$\pm$50 &  3.40$\pm$1.73 & 1.11$\pm$0.28 \\ 
AS2UDSx0065.14 & 34.71820 & $-$4.7114 & 1.21$\pm$0.21 & 3.6 & 34.71819 & $-$4.7114 & 2.35$\pm$0.08 &  3.02$\pm$0.51 &     50$\pm$30 &  1.83$\pm$0.66 & 2.09$\pm$0.40 \\ 
AS2UDSx0094.07 & 34.36970 & $-$5.0719 & 1.59$\pm$0.24 & 3.5 & 34.36974 & $-$5.0719 & 1.85$\pm$0.13 &  2.29$\pm$0.38 &    130$\pm$30 &  2.93$\pm$0.96 & 1.39$\pm$0.21 \\ 
AS2UDSx0124.19 & 34.59010 & $-$5.3770 & 1.24$\pm$0.20 & 3.7 & 34.59008 & $-$5.3769 & 2.29$\pm$0.44 &  1.26$\pm$0.61 &     70$\pm$80 &  1.91$\pm$0.84 & 4.84$\pm$1.00 \\ 
AS2UDSx0131.12 & 34.69560 & $-$5.2402 & 1.28$\pm$0.22 & 3.7 & 34.69562 & $-$5.2401 & 3.38$\pm$0.14 &  0.76$\pm$0.12 &    170$\pm$70 &  1.57$\pm$0.65 & 1.79$\pm$0.22 \\ 
AS2UDSx0131.19$^\ast$ & 34.69380 & $-$5.2417 & 1.36$\pm$0.23 & 3.7 & 34.69379 & $-$5.2416 & 1.03$\pm$1.13 &  0.03$\pm$0.41 &    110$\pm$50 &  3.52$\pm$0.58 & 4.04$\pm$1.34 \\ 
AS2UDSx0131.27 & 34.69531 & $-$5.2435 & 1.71$\pm$0.29 & 3.4 & 34.69528 & $-$5.2434 & 1.56$\pm$0.04 &  2.04$\pm$0.31 &     52$\pm$6 &  3.37$\pm$0.89 & 2.11$\pm$0.11 \\ 
AS2UDSx0143.24 & 34.14653 & $-$5.1127 & 0.56$\pm$0.16 & 3.8 & 34.14657 & $-$5.1126 & 2.52$\pm$0.20 &  2.34$\pm$0.36 &     70$\pm$30 &  0.69$\pm$0.20 & 2.34$\pm$0.42 \\ 
AS2UDSx0145.21 & 34.04314 & $-$4.8945 & 1.40$\pm$0.23 & 3.7 & 34.04321 & $-$4.8944 & 2.15$\pm$0.03 &  1.66$\pm$0.10 &    860$\pm$40 &  2.61$\pm$0.63 & 1.61$\pm$0.01 \\ 
AS2UDSx0148.22 & 34.80909 & $-$5.0150 & 0.28$\pm$0.12 & 3.1 & 34.80904 & $-$5.0149 & 1.85$\pm$0.25 &  0.60$\pm$0.15 &     70$\pm$20 &  0.37$\pm$0.25 & 1.89$\pm$0.44 \\ 
AS2UDSx0152.14 & 34.29503 & $-$5.2110 & 1.27$\pm$0.23 & 3.2 & 34.29502 & $-$5.2109 & 2.48$\pm$0.03 &  0.29$\pm$0.01 &    990$\pm$30 &  1.03$\pm$0.13 & 4.04$\pm$0.01 \\ 
AS2UDSx0159.22 & 34.67400 & $-$5.2268 & 1.15$\pm$0.20 & 3.1 & 34.67390 & $-$5.2270 & 2.94$\pm$0.17 &  0.47$\pm$0.16 &     80$\pm$50 &  1.81$\pm$0.96 & 1.51$\pm$0.26 \\ 
AS2UDSx0188.07 & 34.41810 & $-$5.4740 & 1.56$\pm$0.23 & 3.7 & 34.41810 & $-$5.4740 & 3.25$\pm$0.77 &  0.38$\pm$0.09 &    700$\pm$200 &  1.56$\pm$0.58 & 3.01$\pm$0.91 \\ 
AS2UDSx0197.13 & 34.87302 & $-$4.9771 & 0.70$\pm$0.19 & 3.7 & 34.87315 & $-$4.9772 & 2.19$\pm$0.06 &  8.13$\pm$0.87 &    130$\pm$40 &  1.17$\pm$0.18 & 2.14$\pm$0.16 \\ 
AS2UDSx0200.18 & 34.46520 & $-$5.2519 & 0.97$\pm$0.19 & 3.4 & 34.46538 & $-$5.2520 & 1.83$\pm$0.15 &  5.62$\pm$1.01 &     80$\pm$60 &  1.32$\pm$0.27 & 4.66$\pm$0.44 \\ 
AS2UDSx0202.26 & 34.38260 & $-$4.8398 & 0.57$\pm$0.15 & 3.6 & 34.38248 & $-$4.8398 & 1.64$\pm$0.07 &  0.78$\pm$0.13 &     25$\pm$11 &  1.09$\pm$0.55 & 1.59$\pm$0.18 \\ 
AS2UDSx0223.21 & 34.76431 & $-$5.4481 & 1.65$\pm$0.27 & 3.4 & 34.76419 & $-$5.4482 & 2.44$\pm$0.10 &  0.68$\pm$0.17 &    170$\pm$60 &  3.56$\pm$1.33 & 2.36$\pm$0.07 \\ 
AS2UDSx0258.15 & 34.24234 & $-$5.2806 & 1.19$\pm$0.20 & 3.2 & 34.24219 & $-$5.2806 & 2.29$\pm$0.10 &  0.83$\pm$0.13 &     30$\pm$30 &  2.67$\pm$1.39 & 2.24$\pm$0.51 \\ 
AS2UDSx0280.12 & 34.47139 & $-$5.3614 & 0.96$\pm$0.24 & 3.2 & 34.47130 & $-$5.3614 & 2.67$\pm$1.28 &  0.30$\pm$0.28 &     80$\pm$70 &  1.56$\pm$0.96 & 3.19$\pm$1.39 \\ 
AS2UDSx0280.25 & 34.46905 & $-$5.3633 & 0.40$\pm$0.13 & 3.2 & 34.46907 & $-$5.3632 & 1.61$\pm$0.14 &  0.05$\pm$0.01 &    110$\pm$30 &  0.63$\pm$0.34 & 3.01$\pm$0.01 \\ 
AS2UDSx0285.17 & 34.44131 & $-$4.9120 & 0.81$\pm$0.19 & 3.7 & 34.44136 & $-$4.9120 & 2.08$\pm$0.09 &  1.00$\pm$0.19 &     58$\pm$11 &  1.61$\pm$0.67 & 0.91$\pm$0.14 \\ 
AS2UDSx0289.21 & 34.44750 & $-$4.9426 & 1.55$\pm$0.26 & 3.3 & 34.44760 & $-$4.9428 & 1.53$\pm$0.02 &  1.20$\pm$0.15 &     58$\pm$13 &  3.33$\pm$0.70 & 2.41$\pm$0.09 \\ 
AS2UDSx0301.10$^\dagger$ & 34.46371 & $-$5.1065 & 0.50$\pm$0.16 & 3.1 & 34.46364 & $-$5.1065 & 2.25$\pm$0.96 &  0.21$\pm$0.28 &     20$\pm$30 &  0.74$\pm$0.51 & 3.24$\pm$1.80 \\ 
AS2UDSx0304.13 & 34.85227 & $-$5.4475 & 0.41$\pm$0.13 & 3.2 & 34.85225 & $-$5.4475 & 1.65$\pm$1.07 &  0.08$\pm$0.06 &     20$\pm$30 &  0.63$\pm$0.44 & 1.89$\pm$0.79 \\ 
AS2UDSx0310.20 & 34.47255 & $-$5.1840 & 0.47$\pm$0.15 & 3.5 & 34.47254 & $-$5.1841 & 1.18$\pm$0.46 &  0.72$\pm$0.32 &     40$\pm$30 &  0.64$\pm$0.31 & 5.11$\pm$1.30 \\ 
AS2UDSx0311.09 & 34.46846 & $-$5.1822 & 1.12$\pm$0.19 & 3.2 & 34.46850 & $-$5.1824 & 2.12$\pm$0.15 &  4.27$\pm$0.69 &    140$\pm$40 &  1.63$\pm$0.34 & 3.06$\pm$0.24 \\ 
AS2UDSx0323.17 & 34.43874 & $-$4.8058 & 1.15$\pm$0.20 & 3.8 & 34.43871 & $-$4.8058 & 1.69$\pm$0.05 &  1.91$\pm$0.22 &    250$\pm$40 &  1.44$\pm$0.32 & 1.94$\pm$0.10 \\ 
AS2UDSx0332.17$^\ast$ & 34.53300 & $-$4.8263 & 0.50$\pm$0.16 & 3.3 & 34.53295 & $-$4.8264 & 1.17$\pm$0.09 &  0.24$\pm$0.06 &     60$\pm$13 &  0.66$\pm$0.40 & 1.51$\pm$0.11 \\ 
AS2UDSx0335.15 & 34.88662 & $-$5.2685 & 1.23$\pm$0.21 & 3.6 & 34.88661 & $-$5.2686 & 1.94$\pm$0.07 &  5.62$\pm$0.70 &    220$\pm$30 &  2.05$\pm$0.65 & 1.74$\pm$0.01 \\ 
AS2UDSx0356.11 & 34.11364 & $-$4.8099 & 1.24$\pm$0.20 & 3.3 & 34.11377 & $-$4.8099 & 1.73$\pm$0.01 &  0.50$\pm$0.01 &    252$\pm$3 &  2.03$\pm$0.44 & 2.16$\pm$0.01 \\ 
AS2UDSx0358.15$^\ast$ & 34.36491 & $-$5.4012 & 1.40$\pm$0.22 & 3.9 & 34.36486 & $-$5.4012 & 1.61$\pm$0.58 &  1.02$\pm$0.30 &    640$\pm$170 &  1.87$\pm$0.46 & 3.29$\pm$1.30 \\ 
AS2UDSx0359.13$^\dagger$ & 34.72598 & $-$4.9425 & 1.03$\pm$0.19 & 3.3 & 34.72583 & $-$4.9427 & 2.08$\pm$0.11 &  1.66$\pm$0.21 &     45$\pm$18 &  2.22$\pm$0.85 & 1.39$\pm$0.26 \\ 
AS2UDSx0362.18 & 34.69737 & $-$4.8910 & 0.74$\pm$0.16 & 3.1 & 34.69745 & $-$4.8910 & 3.23$\pm$0.14 &  0.19$\pm$0.04 &     60$\pm$30 &  1.26$\pm$0.73 & 0.99$\pm$0.29 \\ 
AS2UDSx0364.14 & 34.40152 & $-$5.5348 & 1.40$\pm$0.21 & 3.7 & 34.40149 & $-$5.5347 & 1.52$\pm$0.05 &  0.38$\pm$0.03 &    236$\pm$19 &  2.77$\pm$0.52 & 3.29$\pm$0.01 \\ 
AS2UDSx0379.22 & 34.01586 & $-$5.3396 & 0.44$\pm$0.15 & 3.2 & 34.01605 & $-$5.3397 & 1.22$\pm$0.27 &  0.59$\pm$0.32 &     30$\pm$30 &  0.63$\pm$0.34 & 1.44$\pm$0.48 \\ 
AS2UDSx0404.20 & 34.66941 & $-$5.4163 & 1.03$\pm$0.19 & 3.6 & 34.66930 & $-$5.4163 & 2.38$\pm$0.22 &  0.17$\pm$0.14 &    280$\pm$100 &  1.28$\pm$0.47 & 3.29$\pm$0.39 \\ 
AS2UDSx0408.16 & 34.41540 & $-$5.3402 & 0.65$\pm$0.16 & 3.7 & 34.41537 & $-$5.3402 & 1.71$\pm$0.13 &  1.32$\pm$0.03 &    147$\pm$7 &  1.04$\pm$0.39 & 1.89$\pm$0.11 \\ 
AS2UDSx0434.12 & 34.00805 & $-$4.9844 & 1.07$\pm$0.20 & 3.3 & 34.00803 & $-$4.9845 & 3.00$\pm$0.02 &  0.93$\pm$0.02 &    134$\pm$3 &  2.00$\pm$0.91 & 0.69$\pm$0.01 \\ 
AS2UDSx0434.21 & 34.00763 & $-$4.9871 & 2.03$\pm$0.29 & 4.1 & 34.00759 & $-$4.9871 & 2.33$\pm$0.30 &  0.11$\pm$0.05 &     70$\pm$20 &  5.33$\pm$1.54 & 1.91$\pm$0.45 \\ 
AS2UDSx0435.10$^\dagger$ & 34.79616 & $-$4.7379 & 1.35$\pm$0.23 & 3.3 & 34.79638 & $-$4.7378 & 1.74$\pm$0.07 &  0.93$\pm$0.13 &    105$\pm$15 &  2.86$\pm$0.97 & 1.89$\pm$0.01 \\ 
AS2UDSx0435.18 & 34.79638 & $-$4.7391 & 1.85$\pm$0.25 & 4.2 & 34.79636 & $-$4.7391 & 1.70$\pm$0.09 &  3.63$\pm$0.65 &    150$\pm$60 &  2.12$\pm$0.53 & 2.99$\pm$0.32 \\ 
AS2UDSx0443.23 & 34.02782 & $-$4.9518 & 1.07$\pm$0.20 & 3.1 & 34.02771 & $-$4.9519 & 1.38$\pm$0.06 &  4.07$\pm$0.62 &     33$\pm$6 &  2.03$\pm$0.52 & 2.14$\pm$0.11 \\ 
AS2UDSx0448.09 & 34.82097 & $-$4.8737 & 1.00$\pm$0.18 & 3.3 & 34.82096 & $-$4.8736 & 3.40$\pm$1.27 &  1.02$\pm$0.60 &    160$\pm$110 &  1.11$\pm$0.40 & 2.74$\pm$1.74 \\ 
AS2UDSx0464.14 & 34.47000 & $-$5.3954 & 0.95$\pm$0.17 & 3.1 & 34.46987 & $-$5.3954 & 2.44$\pm$0.07 &  1.00$\pm$0.10 &     75$\pm$10 &  2.25$\pm$0.75 & 1.26$\pm$0.04 \\ 
AS2UDSx0469.12 & 34.54256 & $-$5.0194 & 0.72$\pm$0.20 & 3.2 & 34.54246 & $-$5.0194 & 1.67$\pm$0.84 &  0.16$\pm$0.13 &    450$\pm$170 &  0.83$\pm$0.29 & 4.96$\pm$1.21 \\ 
AS2UDSx0469.21 & 34.54182 & $-$5.0218 & 0.52$\pm$0.15 & 3.7 & 34.54176 & $-$5.0218 & 1.50$\pm$0.08 &  1.12$\pm$0.22 &     60$\pm$20 &  1.02$\pm$0.30 & 3.26$\pm$0.26 \\ 
AS2UDSx0472.09$^\dagger$ & 34.76548 & $-$4.7396 & 2.22$\pm$0.31 & 3.6 & 34.76545 & $-$4.7396 & 1.98$\pm$0.13 &  0.83$\pm$0.17 &     80$\pm$30 &  6.19$\pm$1.91 & 3.06$\pm$0.24 \\ 
AS2UDSx0499.22 & 34.60775 & $-$5.5176 & 1.73$\pm$0.24 & 3.8 & 34.60769 & $-$5.5175 & 2.27$\pm$0.08 &  1.26$\pm$0.20 &    130$\pm$100 &  3.10$\pm$1.13 & 2.41$\pm$0.39 \\ 
AS2UDSx0500.22 & 34.53862 & $-$5.2132 & 1.61$\pm$0.28 & 3.2 & 34.53867 & $-$5.2132 & 1.88$\pm$0.18 &  0.02$\pm$0.01 &     16$\pm$11 &  1.31$\pm$1.16 & 1.34$\pm$0.48 \\ 
AS2UDSx0519.12 & 34.42917 & $-$5.4818 & 1.21$\pm$0.22 & 3.5 & 34.42919 & $-$5.4816 & 2.71$\pm$0.33 &  0.35$\pm$0.32 &    600$\pm$300 &  1.22$\pm$0.40 & 3.29$\pm$0.50 \\ 
AS2UDSx0522.17$^{\ast\dagger}$ & 34.77306 & $-$5.0357 & 0.88$\pm$0.17 & 3.3 & 34.77300 & $-$5.0356 & 1.83$\pm$0.41 &  0.17$\pm$0.63 &    570$\pm$180 &  0.95$\pm$0.19 & 4.04$\pm$0.38 \\ 
AS2UDSx0527.17 & 34.87707 & $-$4.8536 & 0.88$\pm$0.17 & 3.3 & 34.87700 & $-$4.8535 & 2.40$\pm$0.19 &  0.89$\pm$0.22 &    560$\pm$140 &  0.95$\pm$0.27 & 3.29$\pm$0.01 \\ 
AS2UDSx0528.13 & 34.05611 & $-$5.0916 & 2.13$\pm$0.26 & 3.9 & 34.05607 & $-$5.0915 & 3.27$\pm$1.07 &  0.62$\pm$0.61 &    400$\pm$300 &  2.93$\pm$1.06 & 2.99$\pm$1.50 \\ 
AS2UDSx0531.24 & 34.80344 & $-$5.4562 & 1.12$\pm$0.20 & 3.3 & 34.80341 & $-$5.4563 & 3.15$\pm$0.55 &  0.15$\pm$0.06 &     70$\pm$50 &  2.38$\pm$1.28 & 1.79$\pm$0.51 \\ 
\hline \noalign {\smallskip}
\end{tabular}
\end{table*}

\newpage

\setcounter{table}{0}

\begin{table*}
\caption{Faint sample catalogue (cont).}
\begin{tabular}{lcccp{0.3cm}ccccccc}
\hline \noalign {\smallskip}
ID & R.A.\ & Dec.\ & $S_{\rm 870\mu m}$ & $\!\!\!$SNR$_{0.5}$ &   R.A.$^{\! \! K}$ & Dec.$^{\!\! K}$ & $z_{\rm med}$ &  $M_{\ast}$ & SFR &  $M_{\rm d}$  & $A_V$ \\
  &  \multicolumn{2}{c}{(J2000)} & (mJy) & & \multicolumn{2}{c}{(J2000)} & & (10$^{11}$\,M$_\odot$) & (M$_\odot$\,yr$^{-1}$) &  (10$^{8}$\,M$_\odot$) &  \\
  \hline \noalign {\smallskip}
AS2UDSx0550.18 & 34.80768 & $-$5.0571 & 1.21$\pm$0.23 & 3.5 & 34.80762 & $-$5.0570 & 1.47$\pm$0.07 &  3.55$\pm$0.50 &     29$\pm$14 &  1.98$\pm$0.72 & 2.06$\pm$0.25 \\ 
AS2UDSx0569.13$^\dagger$ & 34.88377 & $-$4.8918 & 0.98$\pm$0.19 & 3.7 & 34.88374 & $-$4.8918 & 2.23$\pm$0.28 &  1.95$\pm$0.64 &    180$\pm$160 &  1.22$\pm$0.34 & 1.94$\pm$0.76 \\ 
AS2UDSx0575.23$^\dagger$ & 34.67076 & $-$5.4961 & 0.93$\pm$0.22 & 3.9 & 34.67064 & $-$5.4961 & 1.90$\pm$0.22 &  0.62$\pm$0.20 &    142$\pm$11 &  1.45$\pm$0.52 & 2.09$\pm$0.34 \\ 
AS2UDSx0584.12 & 34.36797 & $-$5.3542 & 0.84$\pm$0.18 & 3.4 & 34.36794 & $-$5.3543 & 1.96$\pm$0.15 &  1.51$\pm$0.29 &     59$\pm$17 &  1.22$\pm$0.47 & 2.36$\pm$0.20 \\ 
AS2UDSx0614.18 & 34.51991 & $-$4.7219 & 0.88$\pm$0.19 & 3.8 & 34.51976 & $-$4.7219 & 1.00$\pm$0.01 &  1.41$\pm$0.01 &    465$\pm$1 &  0.52$\pm$0.01 & 3.16$\pm$0.01 \\ 
AS2UDSx0614.24 & 34.52164 & $-$4.7224 & 0.61$\pm$0.16 & 3.3 & 34.52161 & $-$4.7224 & 1.22$\pm$0.02 &  1.74$\pm$0.09 &    142$\pm$7 &  0.79$\pm$0.33 & 2.69$\pm$0.01 \\ 
AS2UDSx0616.24$^\dagger$ & 34.65981 & $-$5.3698 & 0.64$\pm$0.17 & 4.2 & 34.65973 & $-$5.3697 & 1.85$\pm$0.46 &  0.18$\pm$0.15 &    490$\pm$120 &  0.78$\pm$0.17 & 4.49$\pm$0.64 \\ 
AS2UDSx0623.15 & 34.38922 & $-$5.4718 & 0.83$\pm$0.18 & 3.7 & 34.38916 & $-$5.4717 & 3.42$\pm$0.26 &  0.35$\pm$0.08 &     80$\pm$40 &  1.28$\pm$0.75 & 1.46$\pm$0.31 \\ 
AS2UDSx0625.09 & 34.63461 & $-$5.2563 & 1.27$\pm$0.24 & 3.7 & 34.63455 & $-$5.2562 & 1.40$\pm$0.01 &  2.34$\pm$0.03 &    409$\pm$5 &  1.87$\pm$0.78 & 3.19$\pm$0.01 \\ 
AS2UDSx0628.22 & 34.51654 & $-$4.6519 & 0.50$\pm$0.14 & 3.3 & 34.51643 & $-$4.6520 & 3.38$\pm$0.14 &  0.19$\pm$0.05 &     90$\pm$50 &  0.59$\pm$0.28 & 1.16$\pm$0.30 \\ 
AS2UDSx0628.27$^\ast$ & 34.51780 & $-$4.6521 & 0.92$\pm$0.20 & 3.7 & 34.51770 & $-$4.6521 & 3.29$\pm$0.63 &  0.44$\pm$0.20 &    110$\pm$80 &  1.22$\pm$0.65 & 2.06$\pm$0.52 \\ 
AS2UDSx0632.14 & 34.29833 & $-$4.8611 & 0.83$\pm$0.21 & 3.1 & 34.29826 & $-$4.8611 & 2.25$\pm$0.34 &  0.11$\pm$0.02 &    270$\pm$40 &  1.05$\pm$0.36 & 3.11$\pm$0.55 \\ 
AS2UDSx0635.15 & 34.53827 & $-$4.9789 & 0.50$\pm$0.16 & 3.3 & 34.53824 & $-$4.9788 & 1.85$\pm$0.09 &  2.09$\pm$0.30 &     80$\pm$20 &  0.68$\pm$0.26 & 1.29$\pm$0.20 \\ 
AS2UDSx0636.22$^\dagger$ & 34.58341 & $-$4.8416 & 0.73$\pm$0.17 & 3.5 & 34.58338 & $-$4.8415 & 2.23$\pm$0.24 &  1.23$\pm$0.20 &    150$\pm$100 &  0.94$\pm$0.27 & 2.24$\pm$0.66 \\ 
AS2UDSx0641.22$^\dagger$ & 34.41156 & $-$5.3981 & 0.61$\pm$0.15 & 3.4 & 34.41153 & $-$5.3980 & 1.73$\pm$0.44 &  0.81$\pm$0.71 &    360$\pm$70 &  0.79$\pm$0.18 & 2.16$\pm$0.50 \\ 
AS2UDSx0649.16 & 34.38202 & $-$4.9892 & 1.23$\pm$0.21 & 3.9 & 34.38201 & $-$4.9892 & 3.00$\pm$0.21 &  0.41$\pm$0.13 &    90$\pm$50 &  2.47$\pm$1.15 & 1.21$\pm$0.38 \\ 
AS2UDSx0672.16 & 34.73324 & $-$5.0068 & 0.82$\pm$0.18 & 3.1 & 34.73317 & $-$5.0067 & 1.30$\pm$0.08 &  0.87$\pm$0.17 &     75$\pm$19 &  1.73$\pm$0.68 & 1.96$\pm$0.15 \\ 
AS2UDSx0676.22 & 34.08559 & $-$5.2268 & 0.97$\pm$0.18 & 3.7 & 34.08559 & $-$5.2267 & 1.65$\pm$0.11 &  0.52$\pm$0.09 &    170$\pm$60 &  1.57$\pm$0.29 & 3.51$\pm$0.20 \\ 
AS2UDSx0677.18 & 34.24986 & $-$5.0315 & 0.43$\pm$0.14 & 3.7 & 34.24990 & $-$5.0315 & 2.96$\pm$0.11 &  0.72$\pm$0.08 &    104$\pm$11 &  0.41$\pm$0.33 & 0.69$\pm$0.01 \\ 
AS2UDSx0683.08 & 34.15310 & $-$5.5195 & 0.76$\pm$0.17 & 3.2 & 34.15313 & $-$5.5194 & 1.62$\pm$0.09 &  2.14$\pm$0.43 &     90$\pm$50 &  1.02$\pm$0.29 & 1.86$\pm$0.21 \\ 
AS2UDSx0691.18 & 34.10900 & $-$5.4777 & 0.81$\pm$0.20 & 3.2 & 34.10907 & $-$5.4777 & 1.25$\pm$0.29 &  1.26$\pm$0.67 &     50$\pm$60 &  1.22$\pm$0.55 & 1.94$\pm$0.59 \\ 
AS2UDSx0691.20 & 34.11077 & $-$5.4782 & 0.70$\pm$0.17 & 3.8 & 34.11074 & $-$5.4782 & 1.75$\pm$0.61 &  0.87$\pm$0.45 &     30$\pm$40 &  1.13$\pm$0.58 & 3.26$\pm$1.34 \\ 
AS2UDSx0695.16$^\dagger$ & 34.84034 & $-$5.4549 & 0.28$\pm$0.12 & 3.2 & 34.84019 & $-$5.4550 & 2.31$\pm$0.19 &  0.45$\pm$0.10 &     40$\pm$60 &  0.40$\pm$0.23 & 0.79$\pm$0.44 \\ 
AS2UDSx0704.15 & 34.62659 & $-$5.4364 & 0.40$\pm$0.13 & 3.2 & 34.62651 & $-$5.4364 & 2.58$\pm$0.09 &  0.52$\pm$0.10 &     23$\pm$9 &  0.68$\pm$0.43 & 0.86$\pm$0.24 \\ 
AS2UDSx0712.21 & 34.59497 & $-$5.4120 & 1.12$\pm$0.25 & 3.4 & 34.59489 & $-$5.4119 & 1.68$\pm$0.04 &  1.02$\pm$0.16 &    130$\pm$30 &  2.10$\pm$0.44 & 2.09$\pm$0.12 \\ 
\hline \noalign {\smallskip}
\end{tabular}
\end{table*}

\end{document}